\documentclass[%
 reprint,
 showkeys,
 amsmath,amssymb,
 aps,
]{revtex4-2}

\usepackage{graphicx}% Include figure files
\usepackage{dcolumn}% Align table columns on decimal point
\usepackage{bm}% bold math
\begin{document}

\preprint{APS/123-QED}

\title{Thermodynamics of a Single Mesoscopic Phononic Mode}
%\thanks{A footnote to the article title}%

\author{Ilya Golokolenov$^1$}
\author{Arpit Ranadive$^1$}
\author{Luca Planat$^{1,2}$}
\author{Martina Esposito$^1$}
\author{Nicolas Roch$^1$}
\author{Xin Zhou$^3$}
\author{Andrew Fefferman$^1$}
\author{Eddy Collin$^{1,}$}
 \email{eddy.collin@neel.cnrs.fr}
\affiliation{$^1$ Institut N\'eel UGA - CNRS, \\ 25 rue des Martyrs, 38042 Grenoble, France}
\affiliation{$^2$ Silent Waves, 25 Rue Ponsard, 38100 Grenoble, France\\}
\affiliation{$^3$ IEMN, U. Lille - CNRS, \\ av. Henri Poincar\'e, 59650 Villeneuve d'Ascq, France}

\date{\today}% It is always \today, today,
             %  but any date may be explicitly specified

\begin{abstract}
In recent decades, the laws of thermodynamics have been pushed 
down to smaller and smaller scales, within the theoretical field of stochastic thermodynamics and 
state-of-art experiments performed on microfabricated mesoscopic systems. %But these measurements concern mostly thermal properties of electrons and photons. %%% Change
These measurements concern thermal properties of electrons, photons,  and mesoscopic mechanical objects.
Here we report on the measurements of thermal fluctuations of a {\it single mechanical mode} in-equilibrium with a heat reservoir.
The device under study is a nanomechanical beam with a first flexural mode resonating at $3.8~$MHz, cooled down to temperatures in the range from $100~$mK to $400~$mK.
The technique is constructed around a microwave opto-mechanical setup using a cryogenic High Electron Mobility Transistor (HEMT), and is based on {\it two parametric amplifications} implemented in series: an in-built opto-mechanical 'blue-detuned' pumping plus a Traveling Wave Parametric Amplifier (TWPA) stage. 
We demonstrate our ability to resolve energy fluctuations of the mechanical mode in real-time up to the fastest relevant speed given by the mechanical relaxation rate. The energy probability distribution is then exponential, matching the expected Boltzmann distribution. % changed 'law' to 'distribution', Boltzmann law is about black body radiation
The variance of fluctuations is found to be $(k_B T)^2$ with no free parameters. Our microwave detection floor is about 3 Standard Quantum Limit (SQL) at $6~$GHz; the resolution of our fastest acquisition tracks reached about $100~$phonons, and is directly related to the rather poor opto-mechanical coupling of the device ($g_0/2\pi \approx 0.5~$Hz). This result is deeply in the classical regime, but shall be extended to the quantum case in the future with systems presenting a much larger $g_0$ (up to $2\pi \times 250~$Hz), potentially reaching the resolution of a {\it single} mechanical quantum.
We believe that it will open a new experimental field: {\it phonon-based quantum stochastic thermodynamics}, with fundamental implications for quantum heat transport and macroscopic mechanical quantum coherence.
\end{abstract}

%Use showkeys class option if keyword
%display desired
\maketitle

%\tableofcontents
\section{Introduction}
Statistical physics, and by induction thermodynamics, are the basis of our understanding of macroscopic properties from the microscopic entities and laws that structure matter. One of the key results is the second law of thermodynamics, which explains the arrow of time from purely reversible microscopic processes \cite{statbook}.
Fluctuations $\delta X$ of a quantity $X$
are then Gaussian and vanishingly small, leading to a well-defined mean value $\langle X \rangle$.

But many of our intuitive understandings break down at small scales: fluctuations can become as large as mean values, and a specific class of theories known as {\it fluctuation theorems} has been developed to describe them \cite{jarzynski,jarzynskiII}.
With today's technologies, these concepts (and their related paradoxes) can even be probed experimentally using mesoscale and nanoscale devices.
For instance, a {\it Maxwell demon} has been realised by monitoring the charge in a Single Electron Box (SEB), and feeding back this information through a gate voltage controlled by a computer; work is thus extracted \cite{demon}. Such electronic systems are extremely promising, since one can cool them down low enough (tens of milli-Kelvin) so that they behave according to the laws of quantum mechanics. It should then (at least in principle) be possible to probe the impact of quantum coherence on thermodynamic concepts, which is the new exciting field of {\it quantum thermodynamics}
\cite{quantumelec,quantumthermtheory}. 

Beyond electric circuits, thermodynamics is conveying
concepts which are at the intersection of physics, chemistry and biology: and after all, {\it motion} is a key ingredient there. Indeed,  the {\it Landauer erasure principle} has for instance been tested using soft cantilevers and trapped colloids \cite{ciliberto,bellon}, demonstrating that erasing one bit of information produces a minimum $k_B T \, \ln(2)$ amount of heat. Similar stochastic thermodynamics implementations have been realized on DNA molecules, e.g. monitoring their folding/unfolding and extracting work from it \cite{DNA}.
Motion is thus at the core of the definition of {\it heat}: after all {\it phonons} are elementary (quasi-)particles constructed from the (quantized) collective motion of atoms \cite{clelandbook}.
The quantum limit of heat fluctuations \cite{heatfluctu}
and phonon thermal conductance \cite{thermalC,thermalC2} are still  subjects of debate today \cite{pekola,AVSQ}, with very few experiments available in the literature \cite{schwab,olive}.
Besides, centre-of-mass motion of mesoscopic objects is thought to be sensitive to {\it quantum aspects of gravity} (or any other fluctuating fields  
that might be at the source of wave-function collapse) \cite{bassi}.
Such mechanisms predict an imprint on mechanical fluctuations that might be measurable \cite{diosi, vinante}.
But of course, having a large mechanical object cold enough to host very few thermal excitations (population $n_{th} < 1$), while being in-equilibrium with a heat reservoir, is a technological challenge. This has been recently demonstrated with a $15~\mu$m aluminum drumhead device cooled down to $500~\mu$K \cite{Dylan}.

\begin{figure}
        \includegraphics[width=\linewidth]{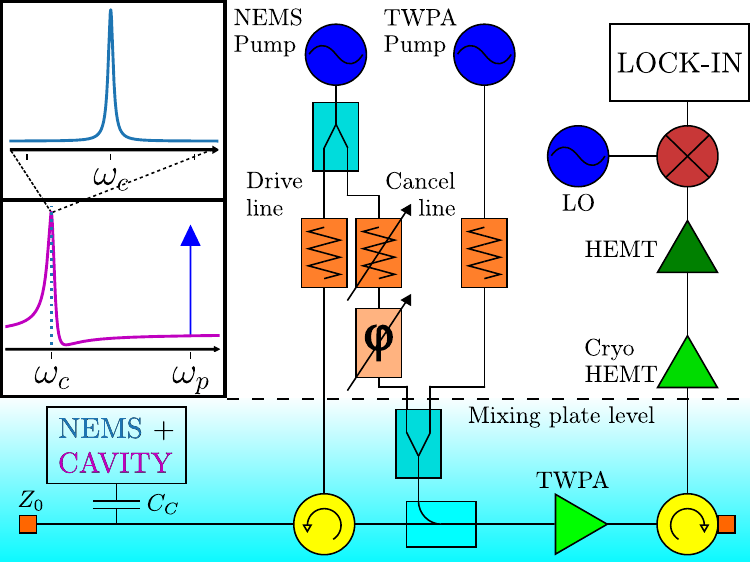}
         \caption{{\bf Main:} Experimental opto-mechanical setup. A microwave pump tone drives the mechanical mode, while the sideband signal is amplified by a TWPA, a cryogenic HEMT plus a room temperature HEMT. The TWPA is powered by a separate pump tone, and is protected from saturation due to the opto-mechanical drive by means of a cancellation tone. The measurement is performed through down-mixing followed by a lock-in amplifier (see text for details). 
        \newline
        {\bf Insets:} Bottom: measurement scheme. A pump tone (of power $P_{in}$) is applied at $\omega_p=\omega_c+\Omega_m$ ('blue pumping' arrow), with $\omega_c$ the cavity resonance frequency (whose susceptibility is shown in violet, arb. units). %That might be excessive due to the absence of y axis ; %%% One needs then to state it!
        Top: mechanical sideband spectrum measured at $\omega_c$, imprinted by the amplified Brownian NEMS motion at $\Omega_m$ (Lorentz curve of half-height-width $\Gamma_{ef\!f}$, arb. height, see text).}
        \label{scheme}
\end{figure}

Studying quantum fluctuations 
%%%%
at equilibrium 
%%%% corr Nico
of a macroscopic mechanical object down to the single quantum might thus be within reach in the near future \cite{AVSQ}. We present in this article a specific scheme enabling such kinds of measurements, based on microwave opto-mechanics. We focus here on a strict {\it classical} description of the experiment, which is a mandatory preliminary step towards the quantum realisation, which we discuss in Conclusion of the present manuscript.
In Section \ref{setup} we describe the apparatus around which the experiment is constructed. In Section \ref{signal} we present the electric circuit modeling leading to the detected signal definition, while in Section \ref{method} the measurement protocol is explained and mathematically analysed. The results are finally discussed in Section \ref{results}.
We separate what is directly the expression of expected properties of a single phononic mode in contact with a heat bath, from features (certainly material-dependent) which are {\it not} expected. The former is an energy power spectrum typical of an {\it Ornstein-Uhlenbeck} process \cite{statbook,ornstein}, with exponentially distributed fluctuations. The latter are visible as $1/f$-type contributions to the spectra and out-of-equilibrium signatures, which will be discussed in the framework of the {\it Two Level Systems} (TLSs) theory \cite{philips,anderson} (Section \ref{oneovf}).

\section{Experimental setup}
\label{setup} 

The opto-mechanical device we use has been presented in Ref.
\cite{XinPRAppl}. It consists of a $50~\mu$m long beam of $300~$nm width and about $100~$nm thickness, embedded in a microwave cavity (gap about $100~$nm). The beam is a bilayer, made of high-stress Silicon-Nitride (SiN) covered with a thin layer of Aluminum. 
The cavity is patterned with a $100~$nm layer of Niobium. The first in-plane flexural mode of the beam resonates at approximately $\Omega_m/(2\pi) = 3.8~$MHz, while the cavity resonates at $\omega_c/(2\pi) = 5.988~$GHz. The motion $x$ of the beam modulates the cavity's  mode effective capacitance $C(x)$, leading to a frequency  change characterized by the (first order) coupling strength $G=d \omega_c/dx$, which is measured to be about $G/(2 \pi) \approx 1.8 \cdot 10^{13} ~$Hz/m. 
The cavity is coupled evanescently (with an effective capacitance $C_c$) to a transmission line which enables to connect the device to the drive/measurement circuitry.

A schematic of the setup is presented in Fig. \ref{scheme}.
A first microwave pump tone is used to drive the opto-mechanics. This signal is split in order to create a 'cancellation line' that opposes whatever remains from this pump at the input of the detection amplifying stage. This cancellation is performed by a computer that checks periodically the pump amplitude on the output, and adjusts a
 voltage-controlled attenuator and phase shifter on the cancellation line. We can suppress this signal by at least $60~$dB. Three distinct microwave amplifiers are in use: first a Traveling Wave Parametric Amplifier (TWPA) that is powered by a separate pump tone \cite{luca}. The characteristics of this device are explicitly given in Appendix \ref{twpagain}, and lead to an equivalent noise at its input of $0.8~$K ($\pm 0.1~$K) for the whole chain (noise figures being quoted at the readout frequency).
 %%%% Corr Luca
 Cancellation and TWPA pump tones are added to the signal line by means of a power combiner and a directional coupler (light blue rectangles). 
The ambient noise coming from the pumps' injection lines is attenuated by about $50~$dB (orange zigzag blocks in Fig. \ref{scheme}). 
Besides, both the NEMS cell and the TWPA are protected by (two-stage) circulators (yellow disks; the orange squares are $Z_0 = 50~\Omega$ loads). The signal is then further amplified by a cryogenic High Electron Mobility Transistor (HEMT)  amplifier from LNF\textsuperscript{\textregistered} with noise temperature $2.5~$K, and then a room temperature HEMT. We finally mix down the signal with a Local Oscillator (LO, a microwave tone shifted from $\omega_c$ by a fixed $\Delta \omega$ frequency) and a Marki\textsuperscript{\textregistered} mixer (brown circle). This megahertz signal at $\Delta \omega$ is finally fed into a ZI\textsuperscript{\textregistered} high frequency lock-in amplifier that is used to demodulate and digitize the data.

The setup is mounted on a commercial BlueFors\textsuperscript{\textregistered} cryostat, and experiments are performed while regulating the mixing chamber base plate from $100~$mK to $400~$mK. At higher temperatures, the TWPA amplifier stops working properly, while at lower temperatures an internal opto-mechanical instability known in the community as 'spikes' corrupts the results \cite{XinPRAppl}. These points shall be commented in more details in Appendix \ref{twpagain} and \ref{spikes} respectively.

\begin{figure}
    \centering
    \includegraphics[width=\linewidth]{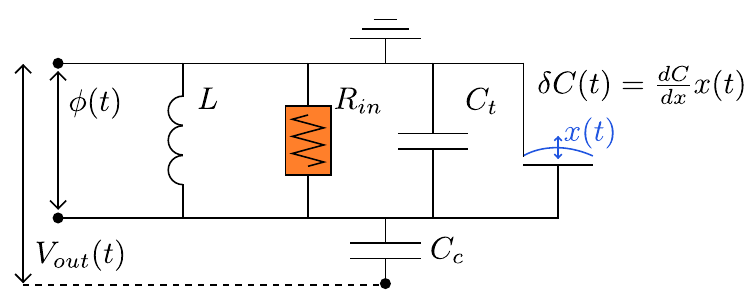}
    \caption{Electric circuit model describing the microwave cavity with a movable capacitor (the boxed 'NEMS+CAVITY' in Fig. \ref{scheme}, see text). }
    \label{electric_circuit}
\end{figure}

\section{Detected signal}
\label{signal}

Our detection scheme is based on 'blue detuned' opto-mechanical pumping (bottom inset Fig. \ref{scheme}) \cite{AKM,XinPRAppl}. 
A microwave pump tone (of power $P_{in}$) is applied at frequency $\omega_p=\omega_c+\Omega_m$. 
The (dynamical) back-action of the light field onto the mechanics leads to {\it anti-damping}, transforming the mechanical relaxation rate $\Gamma_m$ into $\Gamma_{ef\!f}$ (see Appendix \ref{gammaoptic} for details):
\begin{eqnarray}
\Gamma_{ef\!f} & = & \Gamma_m + \Gamma_{opt} , \label{gammaeff} \\
\Gamma_{opt} & = & -  2 G^2 \left(\frac{\kappa_{ext}/2}{\kappa_{tot}}\right) \frac{P_{in}}{\Omega_m \omega_c \, k_0}, \label{gammaopt} 
\end{eqnarray}
with $\kappa_{ext}$ the coupling rate of the cavity to the transmission line, and $\kappa_{tot}=\kappa_{ext}+\kappa_{in}$ the total cavity relaxation rate ($\kappa_{in}$ stands for internal losses).
Since $\Gamma_{ef\!f}<\Gamma_m$, the motion is (parametrically) amplified.
We fit about $100~$kHz for $\kappa_{ext}$ and $200 - 150 ~$kHz for $\kappa_{tot}$ (decreasing with increasing $P_{in}$) on the susceptibility curves (Fig. \ref{scheme} bottom inset, violet line) \cite{XinPRAppl}. 
This decrease of losses with microwave power is presumably due to the presence of Two Level Systems (TLSs), see discussion thereafter.
Note the factor $1/2$ that arises in Eq. (\ref{gammaopt}) because of evanescent coupling.
We write $k_0$ and $m_0$ for the effective spring constant and mass of the mechanical mode respectively, having $\Omega_m^2=k_0/m_0$. 

In the following, we describe the measurement in classical terms, following the electric circuit modeling of Refs. \cite{xintheory,NJPIlya}; only when appropriate shall we make the link to quantum theory.
%%%% add
For the modeling, we shall assume $ \omega \sim \omega_c$, and $Z_0 C_c \omega_c  \ll 1$.
%%%%
The voltage $V_{out}$ to be detected writes:
\begin{equation}
V_{out}(t) = - \omega_c^2 C_c \frac{Z_0}{2} \phi (t) + V_{amp} (t) ,
\end{equation}
with $V_{amp}$ the voltage noise of the detection port, and $\phi$  the generalized flux (primitive integral %primitive 
of the voltage drop) across the microwave  cavity. This cavity is modeled as an $RLC$ circuit, with $C_t=C(0)$ the effective capacitance of the mode in the absence of motion and $L$ its effective inductance, such that $\omega_c^2=1/(L C_t)$ [see Fig. \ref{electric_circuit}]. The microwave damping is modeled through two 
%%%
effective 
%%%
resistors in parallel, an internal $R_{in}$ for all material-dependent losses and a $R_{ext}$ corresponding to the leakage towards the outside (beyond capacitance $C_c$, not shown). %%%% modif Andrew
The total losses simply write $1/R_t=1/R_{in}+1/R_{ext}$, with $\kappa_{in}=1/(R_{in} C_t)$ and $\kappa_{ext}=1/(R_{ext} C_t)$. One defines:
\begin{equation}
 \kappa_{ext} = \omega_c^2  \frac{Z_0}{2} \frac{C_c^2}{C_t},
\end{equation} 
from circuit theory, 
%%%
 considering that the end load of the output line is equal to $Z_0$
%%%
 \cite{xintheory}.
The linear coupling between mechanics and microwaves generates a comb in the output signal, that presents components at $\omega_n = \omega_c + n \, \Omega_m$ with $n \in \mathbb{Z}$. This leads to the expressions \cite{xintheory}:
\begin{eqnarray}
V_{out} (t) & = & \sum_{n=-\infty}^{\infty} \frac{V_{M,n}(t)\, e^{-i \omega_n t} + V_{M,n}(t)^*e^{+i \omega_n t}}{2} , \\
 \phi(t) &= & \sum_{n=-\infty}^{\infty} \frac{\mu_n(t) \,e^{-i \omega_n t} + \mu_n(t)^*e^{+i \omega_n t}}{2} , \\
 V_{amp} (t) & =& \sum_{n=-\infty}^{\infty} \frac{V_{N,n}(t) \, e^{-i \omega_n t} + V_{N,n}(t)^*e^{+i \omega_n t}}{2} ,
\end{eqnarray}
with $V_{M,n}$, $\mu_n$ and $V_{N,n}$
the respective (complex) amplitudes in the frames rotating at $\omega_n$. The motion $x$ is itself expressed in a frame rotating at $\Omega_m$:
\begin{equation}
 x (t) = \frac{x_0(t) \, e^{-i \Omega_m t} + x_0(t)^*e^{+i \Omega_m t}}{2}    , \label{motion}
\end{equation}
with $x_0$ the complex motion amplitude. The drive voltage $V_d$ created by the microwave generator can be defined as:
\begin{eqnarray}
V_d (t) & = & \frac{1}{2} V_p \, e^{-i \omega_p t} + \frac{1}{2} V_p^* e^{+i \omega_p t} \nonumber \\
& + &\sum_{n=-\infty}^{\infty} \frac{V_{P,n}(t) \, e^{-i \omega_n t} + V_{P,n}(t)^*e^{+i \omega_n t}}{2} ,
\end{eqnarray}
with $V_{P,n}$ the complex noise amplitude of the drive field around frequency $\omega_n$. The injected power is thus $P_{in} = |V_p|^2/(2Z_0) $.

In the sideband-resolved limit $\Omega_m \gg \kappa_{tot}/2$, only three components of the comb are relevant: the pump tone at $\omega_p$, and the two sidebands at $\omega_p \pm \Omega_m$ (i.e. $n=\pm 1$). For a 'blue detuned' scheme, $\omega_p=\omega_c+\Omega_m$.
In this case, only the sideband at $\omega_p-\Omega_m=\omega_c$ is measurable, the other one being strongly suppressed. The corresponding voltage amplitude $V_{M,n=-1}$ is found to be \cite{xintheory}:
\begin{eqnarray}
    V_{M,-1}(t) & \approx & -G x_0^*(t) \frac{\kappa_{ext}/2}{\kappa_{tot}} \frac{V_p}{\Omega_m} \nonumber \\
    &+& \frac{\kappa_{ext}}{\kappa_{tot}} V_{P,-1}(t) + V_{N,-1}(t), \label{measure}
\end{eqnarray}
keeping only noise terms inside the cavity.
The reverse scheme is 'red detuned' pumping (pump tone applied at $\omega_c-\Omega_m$), where the sign of $\Gamma_{opt}$ in Eq. (\ref{gammaopt}) is opposite ($-$ should read $+$, leading to {\it attenuation} instead of amplification). The measurable sideband is again the one at $\omega_c$, but it corresponds now to $n=+1$. The voltage amplitude $V_{M,n=+1}$ is then similar to Eq. (\ref{measure}), with a change of sign in front of $G$ and a replacement $x_0^* \rightarrow x_0$. The two schemes 'blue' and 'red detuned' are explicitly compared in the following in order to validate the data analysis.

The mixing process can be formally written as:
\begin{equation}
  \alpha \left[V_{out}(t) \times \cos(\omega_d \, t)\right]_{\mathrm{filter}}  = V_{meas}(t) , \label{demod}
\end{equation}
with $\omega_d$ the frequency of the LO in Fig. \ref{scheme}; we define $\omega_n - \omega_d = \Delta \omega$ the demodulation frequency ($n=\pm 1$ depending on the scheme, 'blue' or 'red detuned' pumping). 
In Eq. (\ref{demod}), the term 'filter' means that only the component at $\omega_n - \omega_d$ is processed, while the one at $\omega_n + \omega_d$
is filtered out. The coefficient $\alpha$ conveniently contains all calibration from the detection chain, which is discussed in more details in Appendix \ref{twpagain}.
One subtlety arises concerning the detection noise background: the component at $\omega_n'=\omega_d-\Delta \omega=\omega_n-2 \Delta \omega$ is actually mixed down equally well as $\omega_n$ in this process, and adds up to the initial noise background $V_{N,n}$ appearing in Eq.  (\ref{measure}). %equally well to what with what? Would propose to add few words, like : 'equally well along with omega_n component'
The voltage digitized by the lock-in amplifier therefore reads:
\begin{eqnarray}
&& V_{meas} (t)  =  \nonumber \\ && \frac{\alpha}{2}\left[ -G \,\bar{x}(t) \frac{\kappa_{ext}/2}{\kappa_{tot}} \frac{V_p}{\Omega_m} + \frac{\kappa_{ext}}{\kappa_{tot}} V_P(t) + V_N(t) \right] \! , \,\, \label{downmix}
\end{eqnarray}
having defined for 'blue detuned' pumping:
\begin{eqnarray}
V_P(t) & = & \frac{  V_{P,-1}(t) e^{-i\Delta\omega t} +   V_{P,-1}^*(t) e^{+i\Delta\omega t}}{2} , \label{noise1}\\
V_N(t) & = & \frac{1}{2}[V_{N,-1}(t)+V_{N,n'}^*(t)] e^{-i\Delta\omega t} \nonumber \\
& +& \frac{1}{2} [V_{N,-1}(t)+V_{N,n'}^*(t)]^* e^{+i\Delta\omega t}, \label{amplinoise} \\
\bar{x}(t) & = & \frac{x_0^*(t) e^{-i\Delta\omega t} + x_0(t) e^{+i\Delta\omega t}}{2}, \label{noise3}
\end{eqnarray}
the noise component due to the pump tone, the amplification chain noise background and the 'effective motion' signal respectively. Note that the latter corresponds exactly to Eq. (\ref{motion}) under the replacement 
$\Omega_m \rightarrow -\Delta \omega $. In Eq. (\ref{amplinoise}), $n'$ refers to the component at $\omega_c-2 \Delta \omega$; no such term exists for the input noise, in the limit $\kappa_{tot} \ll \Delta \omega$ which is taken experimentally 
%%%%%
[we chose arbitrarily $\Delta \omega/(2\pi) = \Omega_m/(2\pi) + 2~$MHz, within the lock-in bandwidth]. % or is it 6 MHz??? I read it carefully and I made a mistake, in this terms delta omega = omega m + 2MHz actually even with minus, but that's not important
%%%XXXXX OK! shall we comment why/how in appendix?
A similar writing holds for 'red detuned' pumping. 
%%%
For more details on the classical circuit theory, we refer the interested reader to Ref. \cite{xintheory}. \\

Up to this point, the time-dependent variables introduced above ($V_P$, $V_N$ and $\bar{x}$)
correspond mathematically to {\it one realization of the stochastic processes they correspond to}.
Let us define $V_{meas}(\omega) = \mathcal{FT}[V_{meas}(t)] (\omega)$ the voltage in frequency space, where $\mathcal{FT}[f(t)] 
(\omega) = \int_{-\infty}^{+\infty} f(t) e^{-i\omega t} dt$. 
Using R. Kubo's notations \cite{kubo}, we define the {\it instantaneous} (i.e. before ensemble-averaging) voltage power spectrum as 
$2 \pi \, S_{V_{meas}}(\omega)\,\delta_0(\omega'-\omega) = V_{meas}(\omega) V_{meas}(\omega')^*$ 
[and similar expressions for the constitutive random variables Eqs. (\ref{noise1}-\ref{noise3})], with  $\omega \in ]-\infty; +\infty [$. 
From Eq. (\ref{downmix}), we obtain:
\begin{eqnarray}
 && S_{V_{meas}}(\omega) = \nonumber \\
 && \frac{|\alpha |^2}{4} \left[ G^2 \left(\frac{\kappa_{ext}/2}{\kappa_{tot}}\right)^{\!2} \frac{|V_p|^2}{\Omega_m^2} \,  S_{\bar{x}}(\omega) \right. \nonumber \\
 && \left. + \left(\frac{\kappa_{ext}}{\kappa_{tot}}\right)^{\!2} S_{V_P} (\omega) + S_{V_N}(\omega) + \mbox{'cross-terms'} \right]\!\! , \,\, \label{Vmeas}
\end{eqnarray}
with 'cross-terms' referring to all cross-correlation spectra.
Knowing that input and output noises are uncorrelated, and that correlations between $\bar{x}$ and $V_P, V_N$ (which are responsible for {\it sideband-asymmetry} \cite{NJPIlya}) are negligible here, these terms shall vanish when computing statistical properties in Section \ref{method}. 

The voltage power spectral densities $S_{V_P} $ and $S_{V_N}$ are reasonably flat over the width of the microwave cavity resonance $\kappa_{tot}$: we can therefore treat them as being white. Besides, since the voltage noise amplitudes $V_{N,-1}$ and $V_{N,n'}$ are uncorrelated, and essentially of equal intensity, the power spectral density  $S_{V_N}$ is {\it twice} the level measured before the mixer. This is the price to pay in the down-mixing process (see Appendix \ref{twpagain}). % we may add:'down-mixing process with two-side mixer' or smth, to specify a bit
%%% I think it would be confusing... We can explain more in Appendix if you want?
Dividing Eq. (\ref{Vmeas}) by $Z_0$, one obtains the (double-sided) power spectral density of detected power [in Watt/(Rad/s), therefore Joule].
Further dividing by $\hbar \,\omega_c$ one converts it into a {\it photon flux} power spectral density [in (photons/s)/(Rad/s), therefore photons]:
\begin{eqnarray}
&& S_{\Dot{\varphi}}(\omega) = \nonumber \\ 
&& 2G^2\left(\frac{\kappa_{ext}/2}{\kappa_{tot}}\right)^{\!2} \frac{P_{in}}{\Omega_m \omega_c k_0} S_n(\omega) \nonumber \\
&& + \left[ \left(\frac{\kappa_{ext}}{\kappa_{tot}}\right)^{\!2} S_{in \varphi}(\omega) + S_{out \varphi}(\omega) \right] ,  \label{spectral}
\end{eqnarray}
where we dropped the calibration factor $|\alpha |^2/4$ for simplicity. 
Note that the formula reads the same for 'red detuned' pumping.
Expressing input and output noises
$S_{in \varphi}$ and $S_{out \varphi}$ in terms of photons enables to evaluate the technique for future quantum measurements \cite{devoret}: we reach about 3 photons which is state-of-art \cite{luca}, see discussion in Appendix \ref{twpagain}.
Explicitly:
\begin{equation}
 S_n(\omega) = \frac{k_0 S_{\bar{x}}(\omega)}{\hbar \Omega_m} ,   
\end{equation}
which corresponds to the instantaneous mechanical energy power spectral density (expressed in phonons), peaked around $\Delta \omega$ (instead of $\Omega_m$).
All of these classical spectra are even, therefore experimentally what is presented for a quantity $X$ is the single-sided $2 \, S_X (f > 0)$, with $ f=\omega/(2 \pi)$ in Hz.

\begin{figure}
        \includegraphics[width=\linewidth]{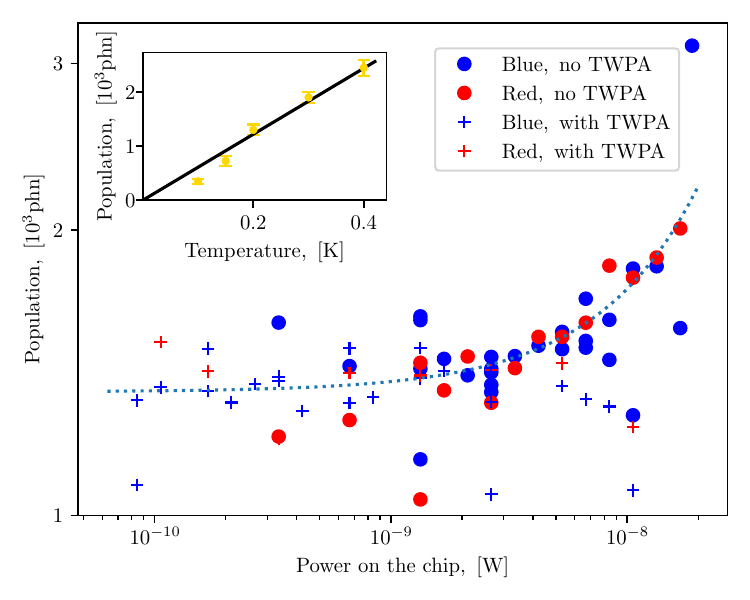}
        \caption{{\bf Main:} Phonon population $\langle E_n \rangle/{\cal G}$ as a function of applied pump power $P_{in}$ measured at $200~$mK. Labels stand for different measuring schemes. The dashed line is a simple guide for the eye (see text; note the log-log scale spanning two orders of magnitude in power).%1.35e3*np.exp(x/4e-8)/1e3 
        \newline
        {\bf Inset:} Intrinsic phonon population (value extrapolated at zero pump power) as a function of temperature. 
%%%
Error bars from the reproducibility scatter of the main graph.        
%%%        
The line is calculated from theory ($k_B T/[\hbar \Omega_m]$, see text).}
        \label{population}
\end{figure}

\section{Measurement protocol}
\label{method}

\begin{figure*}
\includegraphics[width=\linewidth]{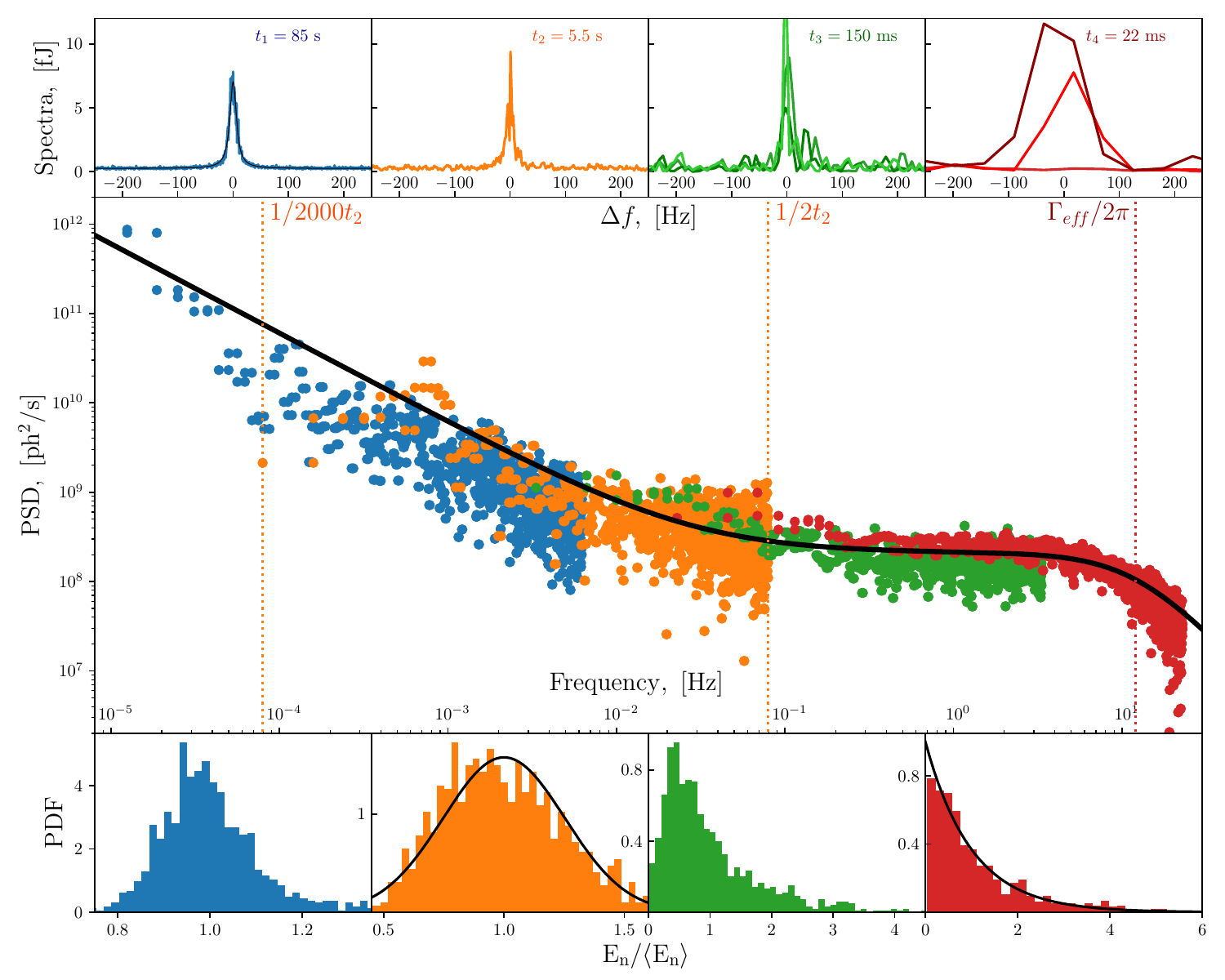}
\caption{ {\bf Main:} The
    central part shows the Power Spectral Density (PSD) of the photon flux $S_{\Dot{\varphi}}$ measured at $T=200 \ \mathrm{mK}$, with a pump power $P_{in}=10~$nW. Different colors correspond to different acquisition speeds ($\delta t$) and durations ($\Delta T= 1000 \, \delta t$), see labels 
    %%%% corr Nico
    $\delta t=\{t_1 , t_2,t_3, t_4\}$ 
    %%%% corr Nico
    top of figure.  
    The black solid line is a fit, which corresponds to the expected spectrum of fluctuations [demonstrating the high-frequency cutoff at $\Gamma_{ef\!f}/(2 \pi)$], but also reveals an  unexpected $1/f$ addendum. Note the overlap between different spectra ('stitching', see text).
    \newline
     {\bf Insets:} Top: typical raw signals for different $\delta t$ (see legend $t_1 - t_4$ and color-code), which are fitted with a Lorentzian function 
%%% add
(black line demonstrated on the left panel)
%%%     
to define sideband peak area $\langle \Dot{E}_{\varphi}\rangle_{\delta t}$, position [shift from reference value $\Delta \omega/(2 \pi)$] and linewidth $\Gamma_{ef\!f}/(2 \pi)$. For slow acquisitions, the peak is well-defined (left graph, see black line fit), while we lose resolution as the acquisition speed is increased; note the different lines plotted in the two right panels, taken from the same respective statistical batches.
     %%% corr Luca
     At the fastest, the   linewidth is essentially given by the sampling (see text).
    Bottom: corresponding Probability Distribution Functions (PDF) versus amplitude normalised to mean (equivalent to $E_n / \langle E_n \rangle$), demonstrating the change of shape as the acquisition speed increases (from Gaussian to Exponential, see black lines and discussion in text).
    }
\label{figbig}
\end{figure*}

In practice, each measured power spectrum is acquired over a finite time $\delta t$ around time $t$: $\langle S_{\Dot{\varphi}} (\omega)\rangle_{\delta t} (t)$. 
Neglecting for now the photon background noise in Eq. (\ref{spectral}), this is simply proportional to the mechanical energy spectrum $\langle S_{n} (\omega)\rangle_{\delta t} (t)$, within a coefficient $\left(\frac{\kappa_{ext}/2}{\kappa_{tot}} \right) |\Gamma_{opt}| \propto P_{in}$.
If the time span $\delta t$ is infinitely long, this quantity should become $t$-independent and reproduce the well known Lorentzian mechanical spectrum, with a half-height-width $\Gamma_{ef\!f}$ and an area $k_B T/(\hbar \Omega_m)$ (schematic in Fig. \ref{scheme}, top inset) \cite{AKM}. This is not perfectly true experimentally because of $1/f$ drifts, see the discussion below.
On the other hand if $\delta t$ becomes infinitely short, one is supposed to obtain (mathematically) a Dirac peak (essentially, the  motion is a well defined oscillation at $\Omega_m$ for timescales $\ll 1/\Gamma_{ef\!f}$), which fluctuates over longer times $t$. Again this suffers from experimental limitations: the frequency resolution is inversely proportional to the acquisition speed, which means that the peak is 'blurred' over a frequency span $1/\delta t$. This aspect is explicitly discussed in Appendix \ref{fastfit}.
The actual experimental dependence of measured spectra for not-too-long, and not-too-short speeds is shown in Fig. \ref{figbig}, top insets.

From the photon flux spectra $\langle S_{\Dot{\varphi}} (\omega)\rangle_{\delta t} $, 
we define the sideband peak power (in photons/s):
\begin{equation}
 \langle \Dot{E}_{\varphi}\rangle_{\delta t} (t)  =  \frac{1}{2 \pi} \int_{-\infty}^{+\infty}   \langle S_{\Dot{\varphi}} (\omega)\rangle_{\delta t} (t)  \,  d\omega   ,
\end{equation}
which is then proportional to the mechanical energy (in phonons):
\begin{equation}
\langle E_{n} \rangle_{\delta t} (t)  =  \frac{1}{2 \pi} \int_{-\infty}^{+\infty}   \langle S_n (\omega) \rangle_{\delta t} (t)  \,  d\omega  .
\end{equation}
Technically, the area of the sideband is not obtained through integration 
%%%% add
(which would be very noisy), but rather with a Lorentz fit 
%%%% add
(from Python\textsuperscript{\textregistered} routines {\it SciPy.optimize.curve\textunderscore fit} \cite{2020SciPy-NMeth} and {\it lmfit.minimize} \cite{newville_matthew_2014_11813}, black line top inset on left, Fig. \ref{figbig}).
%(From Python routines $SciPy.optimize.curve_fit$ \cite{2020SciPy-NMeth} and  $lmfit.minimize$ \cite{newville_matthew_2014_11813}, black line top inset on left, Fig. \ref{figbig}).
%%%%
 Indeed, the background noise that we neglected up to now impacts strongly the quality of the numerical analysis, especially at the fastest acquisition speeds (smallest $\delta t$, see Appendix \ref{fastfit}).
Besides, at not-too-fast speeds, it enables us to fit also both the peak width (extracting thus $\Gamma_m$ from $\Gamma_{ef\!f}$) and the peak position $\Omega_m$ (defined from the demodulation reference). It turns out that these parameters are {\it not} constant, and fluctuate over time; this is explicitly discussed in Section \ref{oneovf}.

Data acquisition is performed over a time $\Delta T$, that we arbitrarily chose as being $1000 \,\delta t$ for convenience (each of our sets is made of $N=1000$ points exactly). 
We construct:
\begin{eqnarray}
 && \langle \Dot{E}_{\varphi} \rangle = \langle\langle \Dot{E}_{\varphi} \rangle_{\delta t}(t) \rangle_{\Delta T}   , \label{means} \\
 && C_{\Dot{E}_{\varphi}}(\tau) = \langle\langle \Dot{E}_{\varphi} \rangle_{\delta t}(t) \, \langle \Dot{E}_{\varphi} \rangle_{\delta t}(t-\tau) \rangle_{\Delta T}  , \label{autocorr}
\end{eqnarray}
 the photon flux mean value and the corresponding auto-correlation function, respectively. 
 In Eq. (\ref{autocorr}), $\tau$ takes discrete values from $-1000 \, \delta t$ to $+1000\, \delta t$. %isnt it 999 actually?) % Heu No?...
 Similar expressions hold for the mechanical energy with $\langle  E_{n} \rangle$, $C_{E_{n}}(\tau)$. For each set, the whole procedure is repeated from 2 - 100 times (depending on acquisition speed)  %%%% true??? In big fig it's 3-100, in all others 2-100; fastests 100, one before - 10, on before - 3-4, and slowest - 2-4;
 in order to improve the quality of the data and assess the impact of $1/f$ drifts on quoted values (indeed, ideally Eqs. (\ref{means},\ref{autocorr}) should be $t$-independent). As always, taking the experimental averaging over $\Delta T$ for an ensemble average is based on a fundamental hypothesis: {\it Ergodicity}. This assumption is not that straightforward here, precisely because 
of the $1/f$ detected features; this point shall be specifically discussed in the following.

\section{Phonon mean population and power spectral density}
\label{results}

In order to interpret the experiment, we first remind the reader of basic classical statistical physics results \cite{statbook,ornstein}. 
We model the mechanical mode as being in contact with a thermodynamic bath at temperature $T$ (the cryostat), and an optical bath at an {\it effective} temperature $T_{opt}$ \cite{NJPIlya}.
We define for the mechanical energy $E_m=\hbar\, \Omega_m \, E_n$: 
\begin{equation}
    \Delta E = E_m - \langle E_m \rangle ,
\end{equation}
the amplitude of fluctuations around the mean $\langle E_m \rangle$. 
This quantity follows the dynamics equation:
\begin{equation}
    \frac{d \Delta E(t)}{dt} = -\left( \Gamma_m + \Gamma_{opt} \right) \Delta E(t) +R_{m}(t) + R_{opt}(t) , \label{relaxE}
\end{equation}
with $R_m$ and $R_{opt}$ the two random energy flows associated with each bath. They verify:
\begin{eqnarray}
 C_m(\tau) & = &\langle R_m (t)R_m (t-\tau)\rangle \nonumber \\
&=& 2\Gamma_m (k_B\, T)^2 \, \delta_0(\tau) , \\
 C_{opt}(\tau) &=& \langle R_{opt} (t)R_{opt} (t-\tau) \rangle \nonumber \\
&=& 2 | \Gamma_{opt} | (k_B\, T_{opt})^2 \, \delta_0(\tau) , \\
 0 &=& \langle R_{m} (t) R_{opt} (t-\tau)\rangle ,
\end{eqnarray}
%%% add??? dunno, its good
meaning that they have no intrinsic finite correlation time, with
the last line stipulating that the two baths are uncorrelated. 
One obtains:
\begin{eqnarray}
\langle E_m \rangle &=& \frac{\Gamma_m (k_B T)  + |\Gamma_{opt}| (k_B T_{opt})}{\Gamma_m + \Gamma_{opt}}, \label{theorEm} \\
S_{\Delta E}(\omega)& =& \frac{2\left[\Gamma_m(k_B T)^2 +|\Gamma_{opt}|(k_B T_{opt})^2 \right]}
{(\Gamma_m + \Gamma_{opt})^2+\omega^2} , \label{theorSEm}
\end{eqnarray}
for the mean energy and power spectral density.
Consider $\Gamma_{opt}=0$; then one recovers the simple case of a {\it canonical ensemble} with a bath at temperature $T$, a situation which has been studied experimentally with macroscopic objects \cite{canonicalPRL}. Energy fluctuations are Gaussian \cite{calormimetry}, a simple consequence of the {\it central limit theorem} because of the large number of degrees of freedom involved. 
However the {\it single mode} statistics is different: it is described by the Boltzmann distribution $p(E)=e^{-E/(k_B T)}/Z$, with in the classical limit the partition function $Z=k_B T$ (ensuring $\int_0^{+\infty} \!p(E)\,dE=1$). One can recover these results from pure (classical) {\it mechanical} arguments, starting from the fluctuation-dissipation theorem and its associated Langevin force (see Appendix \ref{theoryX}). \\

Consider now $\Gamma_{opt} \neq 0$ but $T_{opt} \approx 0$. Eq. (\ref{theorEm})
reads:
\begin{eqnarray}
\langle E_m \rangle &=&  {\cal G} \, k_B T  , \label{equilT} \\
{\cal G} &=&  \frac{ \Gamma_m }{\Gamma_m + \Gamma_{opt}},  
\end{eqnarray}
with ${\cal G}$ the {\it opto-mechanics gain}. We plot in Fig. \ref{population} the mean population of the mechanical mode, recalculated from the mean measured photon flux 
Eq. (\ref{means}). The measurements have been performed with both 'red detuned' (${\cal G}<1$) and 'blue detuned' (${\cal G}>1$) schemes, with the TWPA amplifier on and off (see labels). All the data are in very good agreement, and the remaining scatter is due to the reproducibility of the measurement ($1/f$ drifts). Note that 
$\langle E_n \rangle$ (which in the quantum language is nothing 
%%% corr misprint Nico
but the mode's thermal population $n_{th}$)  verifies $\langle E_n \rangle \gg 1$: we are deeply in the classical limit at all studied temperatures. 

In Fig. \ref{population}, we see that the measured mean phonon population 
%%%% add values
(normalised to gain ${\cal G}$, ranging from 0.5 to 3) 
%%%% corr added Nico
is actually increasing as we increase the injected microwave power $P_{in}$, a phenomenon known as {\it technical heating} in the community (see e.g. Refs. \cite{Dylan,XinPRAppl}). A very natural guess is to assume that the effective temperature of the optical bath $T_{opt} \neq 0$ increases and becomes relevant,  see Eq. (\ref{theorEm}).
This is actually inconsistent. 
In the classical formalism (with a large enough cavity photon population), $T_{opt}=T_{cav} \, \Omega_m/\omega_c$ from back-action;
we safely neglect sideband asymmetry, which would induce a reversed-in-sign correction for 'red' and 'blue pumping' which is not observed here \cite{NJPIlya}. $T_{cav}$ is the effective temperature of the cavity, which could be due to both a real physical heating of the chip (microwave absorption) or to out-of-equilibrium photons arising from the generator noise (see Appendix \ref{twpagain}). At the highest powers (around $15-20~$nW),   $T_{cav}$ would reach about 300$~$K (about 1000$~$photons) which is unphysical (and obviously not observed when measuring directly the output spectrum of the cavity). We thus have to conclude that this effect has another (unknown) origin, with no clear temperature and power dependencies (the line in Fig.  \ref{population} is above all a guide to the eye, 
%%%%
 here a simple linear law).
%%%%
 In inset of Fig. \ref{population}, we plot the mean population extrapolated at zero power. The line is the theoretical prediction from Eq. (\ref{equilT}) with $\Gamma_{opt}=0$: $k_B T/(\hbar \Omega_m)$. The agreement is fairly good, with a slight (unexplained) deviation at low temperatures. 
%%%%%
Note that the scatter is essentially due to the reproducibility of the measurement, impacted by the 1/f noise.
%%%%% 
 We shall comment on the features which are not understood in the following. 

\begin{figure}
        \includegraphics[width=\linewidth]{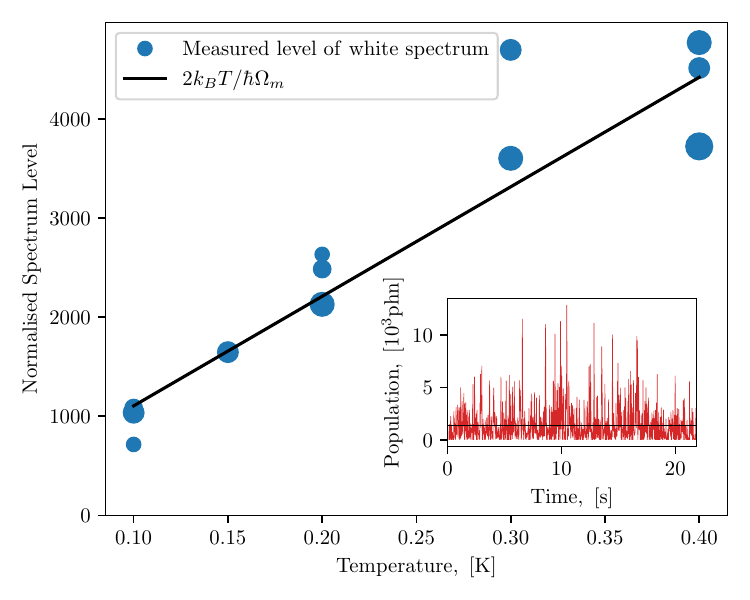}
        \caption{{\bf Main:} Normalised Spectrum level (no units) $ \sqrt{\frac{\Gamma_m \,S_0 }{ {\cal G}^2}} \big / \left(\frac{\kappa_{ext}/2}{\kappa_{tot}} |\Gamma_{opt}|\right)$  at different temperatures and pump powers (size of dots from small/low to big/large $P_{in}$).
        %%% corr ADA
        The line is the theoretical calculation with no free parameters (see text).
        \newline
        {\bf Inset:} Time trace of population fluctuations at the fastest acquisition speed ($200~$mK, power $10~$nW). The line is the mean value. Resolution about 100 phonons, see text for discussion.
        }
        \label{spectlevel}
\end{figure}

From the photon flux correlation function Eq. (\ref{autocorr}), we compute the Fast Fourier Transform ({\it NumPy.correlate} auto-correlation and {\it NumPy.fft} ${\cal FFT}$ algorithms \cite{harris2020array} in Python\textsuperscript{\textregistered}), leading to the experimental spectrum definition:
\begin{equation}
2 S_{\Dot{E}_{\varphi}}(f) = \frac{\delta t}{N^2} \, {\cal FFT} [ C_{\Dot{E}_{\varphi}}(\tau)](f), \label{fft}
\end{equation} % is this correct??? Not exactly, will add corrections; its is Numpy package for Python; the dt works as renormalisation at different acq rates, and division by N^2 makes it kinda unitless after fft, ^2 beacause of autocorrelation.
%%%% OK for me. 
for different acquisition speeds $\delta t$, with the factor 2 on the left-hand-side due to the experimental convention $f>0$. These spectra are plotted in Fig. \ref{figbig}, main graph (see color code for $\delta t$, top insets). The normalisation factor in Eq. (\ref{fft}) takes into account both the number of points $N$ of the discretized acquisition, and the bandwidth $1/\delta t$. For not-too-fast acquisition speeds, the data overlap very well, demonstrating 'stitching'. We conclude that ergodicity is well verified 
%%%% other word?
even at the slowest speeds, where $1/f$ drifts are non-negligible.
However, %%% removed as already pointed out 
the fastest tracks (only red data) 
should be rescaled because of the acquisition finite bandwidth (see Appendix \ref{fastrescale}).

The full spectrum displayed in Fig.  \ref{figbig} is fit by the expression (black full line):
\begin{equation}
   2 S_{\Dot{E}_{\varphi}}(f) = \frac{A_f}{f} + \frac{S_0}{1 + \left(\frac{f}{\Gamma_{ef\!f}/(2 \pi)}\right)^{\!2}}  . \label{fitfunc}
\end{equation}
 The $\Gamma_{ef\!f}$ is actually not fitted, but obtained from the known power dependence of the measured peak width (see Appendix \ref{gammaoptic}), demonstrating very good agreement with data: the mode cannot exchange energy with its environment at speeds exceeding its relaxation rate. 
 The impact of detection noise and of the fitting routine is analysed in Appendix \ref{fastfit}.
 $S_0$ then gives us the level of energy fluctuations while $A_f/f$ corresponds to an unexpected contribution that shall be discussed in the next Section \ref{oneovf}.  

Applying now the assumption $T_{opt} \approx 0$ to Eq. (\ref{theorSEm}), one can write:
\begin{equation}
    S_{\Delta E}(|\omega| \ll \Gamma_{ef\!f}) \approx  {\cal G}^2 \, \frac{2 \, (k_B T)^2 }
{\Gamma_m } ,
\end{equation}
with $2\, (k_B T)^2/\Gamma_m$ the $\omega \rightarrow 0$ value that characterises an {\it unpumped} mode ($\Gamma_{opt}=0$). Reversing this expression and making use of the transduction coefficient between the optical and the mechanical fields, we can therefore recalculate from the best fit value of $S_0$ the actual thermodynamical mechanical fluctuation level. Comparison between 'red detuned' (${\cal G} <1$) and 'blue detuned' (${\cal G} >1$) schemes is discussed in Appendix \ref{compareRB}. Making this experiment at various temperatures, we present this quantity (normalised to the phonon energy $\hbar \Omega_m$) as a function of $T$ in Fig. \ref{spectlevel}. The black line is the theoretical prediction, with no free parameters, demonstrating very good agreement with data.
%%% Add comment noise ADA
The scatter seems to be due to our reproducibility and fitting capability 
%%%% added
(see Appendix F), with no specific link to the microwave power $P_{in}$.
%%%%
It confirms the magnitude of the variance $\langle \Delta E^2 \rangle =1/(2 \pi) \int_{- \infty}^{+ \infty} S_{\Delta E}(\omega) d\omega =(k_B T)^2 $ in this canonical ensemble;
the subtlety being that the specific heat associated with the single-mode is {\it precisely} $k_B$ \cite{canonicalPRL}.

Finally, from the acquired time-tracks of $\langle \Dot{E}_{\varphi}\rangle_{\delta t} (t)$ we can build histograms; this is done for each acquisition speed $\delta t$, see bottom insets in Fig. \ref{figbig}. We plot them with an area normalised to 1 [directly compatible with a Probability Distribution Function (PDF)], and an energy amplitude axis normalised such that the mean is also 1.
At the fastest speed the shape is clearly exponential (right plot; the line is a theoretical function with no fit parameter). The standard deviation is then also 1. However as we slow down the acquisition, the distribution becomes gradually Gaussian (see line middle-left plot of Fig. \ref{figbig} bottom insets), with a {\it smaller} standard deviation, as it should.
%%%% added 
Note that the first PDF (on the left) does not display a fit, the histogram being quite impacted by the slow drifts of the (non-stationary) 1/f noise. \\

\begin{figure}
        \includegraphics[width=\linewidth]{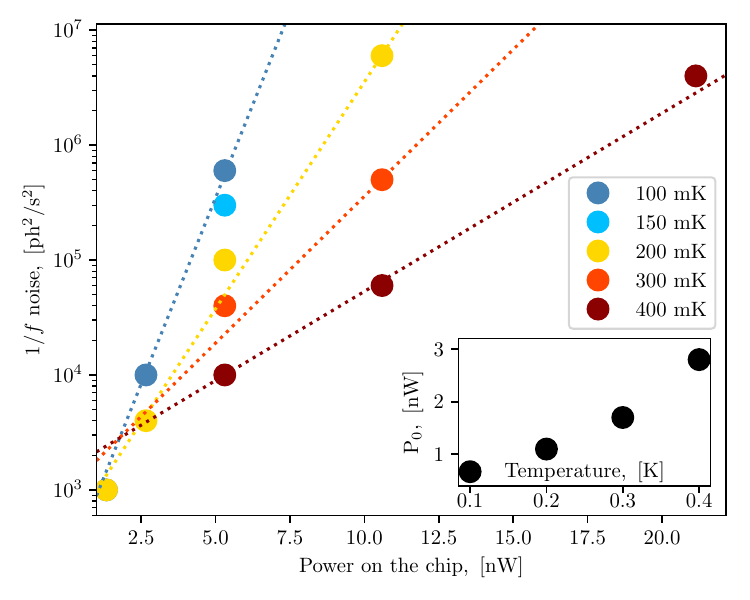}
        \caption{{\bf Main:} $A_f$ constant of $1/f$ noise contribution fitted at different powers and temperatures, in units of photons on chip. Dashed lines are exponential fits: $A_f = A_0(T) e^{P/P_0(T)} $ (see text).
        \newline
        {\bf Inset:} $P_0$ fit parameter from main graph as a function of temperature $T$.}
        \label{1ovf}
\end{figure}

An attempt had been made in Ref. \cite{Dylan} to develop the measurement method producing Fig. \ref{figbig}. However the resolution was very far from enough for this purpose (no TWPA was used), and the 
heavy averaging was essentially filtering out the thermodynamic contribution, leaving only what should have been the equivalent of our $1/f$ term, see discussion in Appendix \ref{sliding}.
On the other hand, we reach here at the fastest tracks a resolution of about 100$~$phonons (see real-time data in inset Fig. \ref{spectlevel}), being limited only by our relatively poor opto-mechanical coupling $G$. In quantum mechanics terms with the zero point fluctuation $x_{zpf} \approx 27~$fm, we have $g_0 = G \, x_{zpf}  \approx 2 \pi \, 0.5~$Rad/s. Using drumhead aluminum devices in the future, one can reach couplings as high as $g_0\approx 2 \pi \, 250~$Rad/s, winning therefore about a factor $\sim 10^5$ on the detected signal (all other parameters being kept equivalent) \cite{NJPIlya}.

\section{$1/f$ noises}
\label{oneovf}

A striking unexpected feature observed in our measurements is the $1/f$ contribution to the photon flux fluctuations, main plot in Fig. \ref{figbig}. We show the fit parameter $A_f$ in Fig. \ref{1ovf} as a function of both injected power $P_{in}$ and temperature $T$. We observe that this coefficient can be fit by an exponential input power dependence (note the $y-$axis of the Figure), with a smooth temperature dependence (see Inset). 

The origin of this effect remains unknown. It is not even clear if it originates in the phonon or in the photon field (see discussion in Appendix \ref{compareRB}, comparing 'red' and 'blue detuned' pumping), which is why we characterize it in terms of photons. Besides, since $1/f$ drifts are responsible for very slow (close to $\omega \rightarrow 0$) dynamics, one could wonder whether this signature has to be linked to the {\it technical heating} of Fig. \ref{population} (a supposedly true D.C. effect). Again, this remains an open question.

\begin{figure}[t]
        \includegraphics[width=\linewidth]{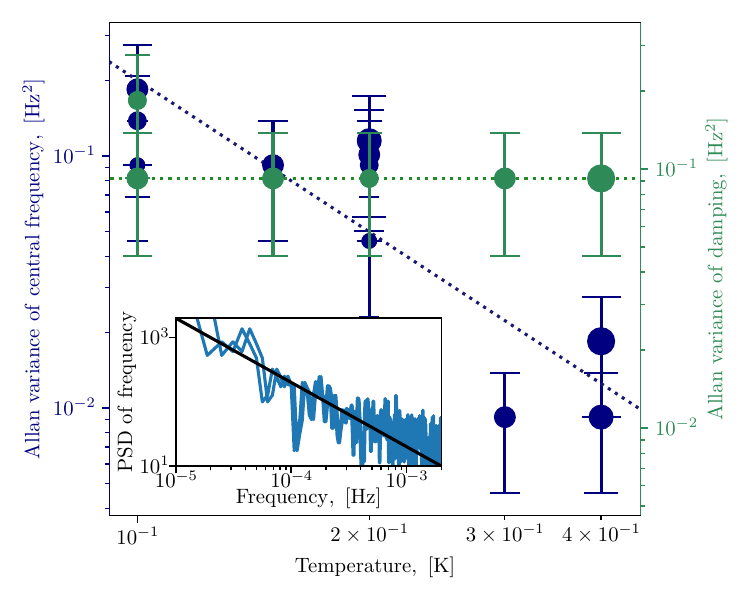}
        \caption{{\bf Main:} Allan variance of mechanical frequency  $\Omega_m/(2 \pi)$ (blue) and of mechanical damping $\Gamma_m/(2 \pi)$ (green) as a function of temperature. Tracks acquired with our slowest acquisition speed, over about 14 h. 
        %%% corr
        Size of dots for different microwave powers (from small/low, to big/large as in Fig. \ref{spectlevel}).
        %($ln(10^{-3}/10^{-5})=ln100=4.6$; corr.to 14h.) 
        Dashed lines are guides for the eyes. %, corresponding to  $0.002T^{-2}$(blue) and constant $0.092$(green).
        \newline
        {\bf Inset:} Typical power spectral density, here computed for frequency (at $200~$mK, with drive power $10~$nW). The line is a $1/f$ fit (see text).
        %%%% add
        The doubling of the line comes from the folding of the negative frequency axis onto the positive one.
        }
        \label{1ovfdampfreq}
\end{figure}

Besides, one should also keep in mind that the mechanical parameters $\Omega_m$ and $\Gamma_m$ are also fluctuating; this had been already reported in Ref. \cite{Dylan} in the framework of low-temperature opto-mechanics, but also in more conventional experiments \cite{SansaNat,oliveACS}. Since we fit the mechanical response (on not-too-fast tracks), we can extract these parameters and compute their statistical properties. This is summarized in Fig. \ref{1ovfdampfreq}.
The probability distributions look reasonably Gaussian, and the power spectral densities present a clear $1/f$ trend
%%%% modif
 (see example in Inset). 
We find out that the damping noise is essentially constant in temperature $T$, which is consistent with findings from Ref. \cite{oliveACS} taken at slightly higher temperatures. On the other hand, the frequency noise {\it grows} as we cool down, a feature also seen in Ref. \cite{Dylan} down to much lower temperatures. 
Both damping and frequency fluctuations are of the same order as reported values for SiN beam devices cooled at cryogenic temperatures \cite{oliveACS}.
%%%% Comment scatter ADA
The scatter in Fig. \ref{1ovfdampfreq} is rather large (as is usually the case when measuring $1/f$), but no specific drive power dependence can be seen.
%%%%
Again, the mechanism behind these features might be linked to the previous properties impacting energy fluctuations, but no microscopic theory has been formulated yet.

It is nonetheless tempting to 
imagine that Two Level Systems (TLSs) are responsible for these facts. Indeed, low temperature mechanical properties of microfabricated structures are  analysed in the framework of this model \cite{TLSdavis,TLSnems}, while experiments on microwave cavities also reported a growing frequency variance with lowering temperature, which was interpreted as a signature of interacting TLSs \cite{TLSsupra}.
Furthermore, an internal instability of microwave opto-mechanics has been reported below typically $150~$mK  \cite{XinPRAppl}. This feature, nicknamed 'spikes', is discussed in Appendix \ref{spikes}. The deviation at low $T$ of the mechanical mode energy with respect to theory (Fig. \ref{population} inset) might be related to this; the fact that the most impacting $1/f$ signatures grow as we cool down is a rather intriguing fact, that might suggest that a profound link exists between them. 

\section{Conclusion}

We report on a technology that enables us to measure in real-time the energy fluctuations of a mesoscopic mechanical mode. 
The setup is constructed around a state-of-art microwave opto-mechanical cryogenic platform, presenting a background noise of 3$~$SQL (3 photons at 6$~$GHz).
The resolution is then about 100$~$phonons at our fastest acquisition rates (about 20 milliseconds), with a 4$~$MHz mode.
The limiting parameters are the intrinsic losses of the TWPA \cite{luca}, and mostly the weak opto-mechanical coupling of the device we used here \cite{XinPRAppl}. We believe that both aspects can be greatly improved, leading to an effective resolution on the detected sideband spectrum equivalent to a {\it single mechanical quantum}.   

This means not only that these experimental capabilities surpass the best microfabricated calorimetric setups to date (zepto-Joule calorimetry \cite{adib}), but also that the {\it phonons themselves} become the quantum bath being probed;
performing quantum calorimetry with phonons, opening thus a new experimental field  \cite{quantCalo,quantCJukka}.

The next experimental step is therefore to mount this measurement setup onto a cryostat enabling to 'brute-force' cool down to the quantum regime such MHz mechanical modes, as demonstrated in Ref. \cite{Dylan}.
{\it Quantum stochastic thermodynamics} experiments would be at reach \cite{AVSQ}, but this requires to further analyse the setup in quantum mechanics terms 
 (the theoretical treatment presented here being purely classical). 

The measurement is constructed around the observable $\hat{x} \propto \hat{a}+\hat{a}^\dag$ (motion amplitude), not energy $\hbar \Omega_m\, \hat{n} \propto \hat{a}^\dag \hat{a}$, which means that we shall not detect single-phonon tunnelings per se, but their (dispersive) imprint onto the optical field. 
This should nonetheless enable to study single phonon events, transposing to mechanics what has been beautifully achieved for electrons. As an example, one should be able to demonstrate, at extremely low temperatures where the mechanical thermal population $n_{th}<1$, how the system can be {\it absolutely free} of excitations over macroscopic timescales (similarly to electrons in a superconductor) \cite{jukkasingle}: a rather counter-intuitive possibility, which essentially means that the system could be said to be a $T=0~$K exactly for a short period of time, while obviously {\it on average} $T>0~$K is always recovered. % Wow

\begin{acknowledgments}

We wish to acknowledge the use of the N\'eel {\it Nanofab} facility.
The Authors are grateful to 
J. L. Garden, O. Bourgeois, O. Maillet, and A. D. Armour for very useful discussions.
 We acknowledge support from the ERC CoG grant ULT-NEMS No. 647917 (E.C.), and StG grant UNIGLASS No. 714692 (A.F.), and by the European Union's Horizon
2020 research and innovation program under grant
agreement no. 899561 (N.R.).
 The research leading to these results has received funding from the European Union's Horizon 2020 Research and Innovation Programme, under grant agreement No. 824109, the European Microkelvin Platform (EMP).
\end{acknowledgments}

\appendix

\section{From Langevin force to Boltzmann distribution}
\label{theoryX}

One can easily construct the energetics description of a mode from its motion, at least in the so-called high-$Q$ limit.
Consider the dynamics equation of a harmonic oscillator ($m_0$ being its mass, and $k_0$ its spring constant, $\Omega_m^2=k_0/m_0$):
\begin{equation}
  \ddot{x} + \Gamma_m \, \dot{x} + \Omega_m^2 \, x =  \delta F/m_0 ,  \label{newton}
\end{equation}
with $\delta F(t)$ the Langevin force linked to the damping $\Gamma_m$ through the {\it fluctuation-dissipation theorem}. Both originate from a thermal bath at temperature $T$, and the stochastic force is by definition described by a centered ($\langle \delta F \rangle=0$) Gaussian probability distribution with correlator:
\begin{eqnarray}
     C_{\delta F}(\tau) &=& \langle \delta F(t) \delta F(t-\tau)\rangle \nonumber \\
&=& 2 m_0 \Gamma_m \, k_B T \, \delta_0(\tau) . \label{dFcorr}
\end{eqnarray}
The presence in the above equation of the Dirac distribution simply means that there is no finite correlation time characterizing the bath (the associated spectrum is white). 
This obviously poses a mathematical problem for our definitions: the variance of this noise which defines the width of the Gaussian probability distribution is infinite (since it is related to the integral of the fluctuation spectrum). It shall not impact the final result of the modeling, which is cut-off at high frequencies by the mechanical relaxation rate. One should therefore clarify that $\delta F$ fluctuations
are Gaussian for any  bandwidth $\Delta \omega$ cut in the white noise spectrum,  around any frequency $\omega_0$.

Let us now transpose the dynamics into the Rotating Frame associated to the mode (at frequency $\Omega_m$):
\begin{eqnarray}
\delta F(t)&=& F_X(t) \cos(\Omega_m t) + F_Y(t) \sin(\Omega_m t) ,  \label{FXFY} \\
x(t) & = & X(t) \cos(\Omega_m t) + Y(t) \sin(\Omega_m t) ,
\end{eqnarray}
having introduced the two quadratures of force and motion. Eq. (\ref{newton}) can be rewritten, in matrix form:
\begin{eqnarray}
    \begin{pmatrix}
    1 & +\frac{1}{2Q} \\
    -\frac{1}{2Q} & 1 
    \end{pmatrix}
    \begin{pmatrix}
    -\dot{X}  \\
     +\dot{Y} 
    \end{pmatrix}
   & =& - \frac{\Gamma_m}{2}
     \begin{pmatrix}
    1 & 0 \\
    0 & 1 
    \end{pmatrix}
    \begin{pmatrix}
    -X  \\
     +Y 
    \end{pmatrix} \nonumber \\
&&+ \frac{\Omega_m}{2 k_0}
    \begin{pmatrix}
    F_Y  \\
     F_X 
    \end{pmatrix} , \label{matrixEq}
\end{eqnarray}
with $Q=\Omega_m/\Gamma_m$ the quality factor, having neglected the slow components $\ddot{X},\ddot{Y} $ (Rotating Wave Approximation, valid for $Q \gg 1$). Similarly, we write:
\begin{eqnarray}
\dot{x}(t) & = & \Omega_m \left[ -X(t) \sin(\Omega_m t) + Y(t) \cos(\Omega_m t) \right] ,
\end{eqnarray}
for the velocity, neglecting $\dot{X},\dot{Y}$. 
From the definitions of kinetic energy $E_c=m_0 \, \dot{x}^2/2$ and potential energy $E_p=k_0\, x^2/2$, we simply obtain for the total energy $E_m=E_c+E_p$:
\begin{equation}
  E_m(t)= k_0 \frac{ X(t)^2+Y(t)^2}{2}.  \label{energyEm}
\end{equation}

Let us take the limit $Q \rightarrow + \infty$ in Eq. (\ref{matrixEq}); the $X$ and $Y$ equations then separate. Multiplying the first one by $k_0 \, X$, and the second one by $k_0 \, Y$ we write:
\begin{equation}
\left\{
\begin{aligned}
k_0 \, X \, \dot{X} & =& - \frac{\Gamma_m}{2} k_0 X^2 -\frac{\Omega_m}{2} X \, F_Y, \\
k_0 \, Y \, \dot{Y} & =& - \frac{\Gamma_m}{2} k_0 Y^2 +\frac{\Omega_m}{2} Y \, F_X,
\end{aligned} 
\right.  
\end{equation}
which after adding-up leads to the result:
\begin{eqnarray}
    \frac{d E_m}{d t}  &=& - \Gamma_m \, E_m  \nonumber \\
    && + \frac{\Omega_m}{2}\left[ Y \,F_X - X \, F_Y \right] . \label{equaEtosolve}
\end{eqnarray}
This equation can be recast into the form of
Eq. (\ref{relaxE}) by introducing the energy difference $\Delta E(t)=E_m(t) - \langle E_m \rangle$ and {\it the bath stochastic energy flow} $R_m(t)$:
\begin{equation}
 \!\!  R_m(t) =  \frac{\Omega_m}{2}\left[ Y(t) F_X(t)- X(t) F_Y(t)\right] - \Gamma_m \langle E_m \rangle,  \label{Rm}
\end{equation}
obtained here in the high-$Q$ limit; a more generic discussion can be found in Ref. \cite{PRELangevin}. 
The mean energy can be inferred from the {\it equipartition result} (see thereafter): $\langle E_m \rangle = k_B T$. 
Eq. (\ref{equaEtosolve}) is finally easily solved in frequency-space as:
\begin{equation}
    S_{\Delta E}(\omega)  =  \frac{ S_{R_m}(\omega) }
{\Gamma_m^2+\omega^2} , \label{Senergy}
\end{equation}
with $ S_{R_m}$ the spectrum associated to $R_m$. We should now construct the statistical properties of this variable, from the initial properties of $\delta F$. \\

To do so, Eq. (\ref{Rm}) is rewritten as: 
\begin{eqnarray}
  R_m(t) & = &  R_X(t)+R_Y(t) - \Gamma_m \langle E_m \rangle, \label{Rm2}  \\
  R_X(t) & = & \frac{\Omega_m}{2}  (\chi * F_X)(t) \, F_X(t), \\
  R_Y(t) & = &\frac{\Omega_m}{2}(\chi * F_Y)(t) \, F_Y(t),
\end{eqnarray}
where we introduced the mechanical susceptibility $\chi(t)$ Fourier Transform (in the rotating frame), and $*$ designates the convolution product [$(f * g)(t) = \int_{-\infty}^{+\infty} f(t-x)g(x) \, dx $]. We have:
\begin{eqnarray}
\chi(t) & = & \frac{\Omega_m}{2k_0} \, e^{-\frac{\Gamma_m}{2} t} \,\, \Theta(t) , \label{xi} \\
\langle F_X(t) F_X(t-\tau) \rangle &=& 4 m_0 \Gamma_m \, k_B T \, \delta_0(\tau) , \label{eq1F} \\
\langle F_Y(t) F_Y(t-\tau) \rangle &=& \langle F_X(t) F_X(t-\tau) \rangle ,  \label{eq2F} \\
\langle F_X(t) F_Y(t-\tau) \rangle &=& 0 , \label{eq3F}
\end{eqnarray}
with $\Theta(t)$ the Heaviside function (0 for $t<0$, and 1 for $t>0$; for the time being, we only require $\Theta(0)$ to be finite). 
The relations Eqs. (\ref{eq2F},\ref{eq3F})
simply state that the phase of the random force is irrelevant, which would not be the case in the presence of a {\it squeezed noise}.
Eq. (\ref{xi}) solves Eq. (\ref{matrixEq}) in the limit $1/Q \approx 0$, while Eq. (\ref{eq1F}) 
is deduced from Eqs. (\ref{dFcorr},\ref{FXFY}) [note the extra factor of 2 in the rotating frame noise amplitude]. $F_X$ and $F_Y$ have by construction the same probability distribution as $\delta F$ (namely Gaussian).
The mean values verify:
\begin{eqnarray}
     \langle R_X \rangle &=&  \langle R_Y \rangle \nonumber \\
   &    = & \Gamma_m  \, k_B T  \!\! \int_{-\infty}^{+\infty} \!\!\!\!\!\!\! e^{-\frac{\Gamma_m}{2} (t-x)} \Theta(t-x) \delta_0(x-t) \,  dx \nonumber \\
   & = &\Gamma_m  \, k_B T \, \Theta(0), 
\end{eqnarray}
which introduces the value $\Theta(0)$ which has not been defined yet. In order to impose $\langle R_m \rangle =0$, we {\it have to} take $\Theta(0)=1/2$.

Consider now the correlation functions of the type:
\begin{eqnarray}
&& \langle R_A(t) R_B(t') \rangle = \left(\frac{\Omega_m^2}{4 k_0} \right)^{\!2} \!\! \times \nonumber \\
&& \int_{-\infty}^{+\infty}  \!\!\! \int_{-\infty}^{+\infty} \!\!\!\!  e^{-\frac{\Gamma_m}{2} (t-x)} \Theta(t-x) \, e^{-\frac{\Gamma_m}{2} (t'-x')} \Theta(t'-x') \nonumber \\
&&  \langle F_A(x) F_A(t) F_B(x') F_B(t') \rangle \,\, dx dx', \label{correlAB}
\end{eqnarray}
with $A,B=X,Y$ in all possible combinations. 
The second order force correlator can be decomposed using Wick's theorem:
\begin{eqnarray}
&& \langle F_A(x) F_A(t) F_B(x') F_B(t') \rangle  = \nonumber \\
&& +\langle F_A(x) F_A(t) \rangle \, \langle F_B(x') F_B(t') \rangle \nonumber \\
 && + \langle F_A(x) F_B(x') \rangle \, \langle F_A(t) F_B(t') \rangle \nonumber \\
 && + \langle F_A(x) F_B(t') \rangle \, \langle F_B(x') F_A(t) \rangle .
\end{eqnarray}
Reinjecting this result into Eq. (\ref{correlAB}), one obtains:
\begin{eqnarray}
&&\!\!\!\!\!\!\!\!\!\!\!\!\!\!\!\!\!\!\!\!\!\!\!\!\!\!\!  \langle R_X(t) R_X(t') \rangle = \langle R_Y(t) R_Y(t') \rangle = \left(\Gamma_m k_B T \right)^2  \Theta(0)^2 \nonumber \\
&& \!\!\!\!\!\!\!\!\!\!\!\!\!\!\!\!\!\!\!\!\!\!\!\!\!\!\! +
\Gamma_m  \left(k_B T \right)^2 \left[ \delta_0(t-t')+\Gamma_m \Theta(t-t')\Theta(t'-t) \right] , \label{XXYY}\\
&& \!\!\!\!\!\!\!\!\!\!\!\!\!\!\!\!\!\!\!\!\!\!\!\!\!\!\!  \langle R_X(t) R_Y(t') \rangle = \langle R_Y(t) R_X(t') \rangle = \left(\Gamma_m k_B T \right)^2  \Theta(0)^2 \! . \,
\end{eqnarray}
In Eq. (\ref{XXYY}), the product $\Theta(t-t')\Theta(t'-t)$ can obviously be dropped when compared to the $\delta_0(t-t')$ term.
Finally, making use of all of these findings,  we deduce from Eq. (\ref{Rm2}):
\begin{equation}
    \langle R_m(t) R_m(t') \rangle = 2 \Gamma_m (k_B T)^2 \delta_0 (t-t'), 
\end{equation}
which leads to the corresponding (white) spectrum $S_{R_m} (\omega) =  2 \Gamma_m (k_B T)^2 $; Eq. (\ref{Senergy}) therefore reproduces Eq. (\ref{theorSEm}), as it should. \\

\begin{figure}[t]
        \includegraphics[width=\linewidth]{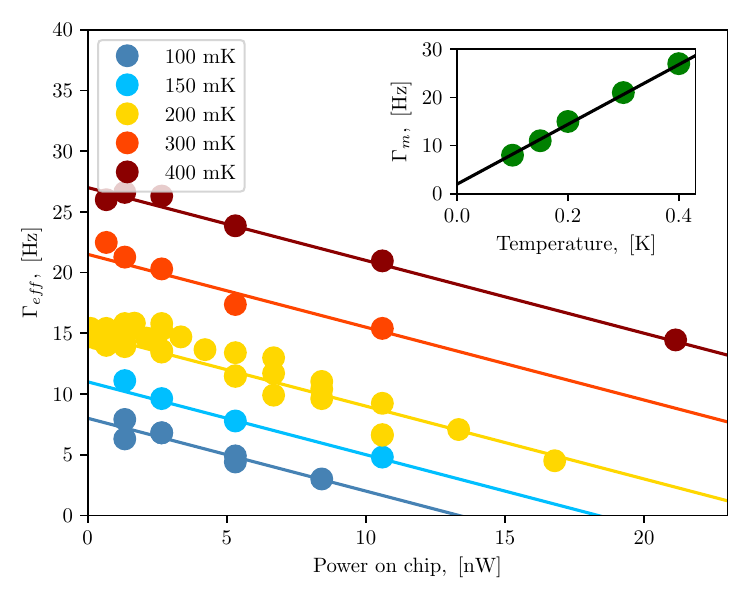}
        \caption{{\bf Main:} Power dependence of the effective damping (peak half-height-width) $\Gamma_{ef\!f}$, at different temperatures. The lines are fits leading to $\Gamma_m$ ($P_{in}=0$ values) and coupling $G$ (slope), see text.
        \newline
        {\bf Inset:} 
        Temperature dependence of mechanical damping $\Gamma_m$. The line is a linear fit \cite{XinPRAppl}.
        }
        \label{fig_damping}
\end{figure}

To conclude this Appendix, let us focus on   the distributions of the random variables. Since the forces $F_X, F_Y$ are Gaussian distributed, their corresponding motions $X, Y$ are also Gaussian (linear response). Their Probability Distribution Functions (PDF) write:
\begin{eqnarray}
p(X) &=& \frac{1}{\sqrt{2 \pi \, \sigma_X^2}}e^{-\frac{X^2}{2 \sigma_X^2}}, \\
p(Y) &=& \frac{1}{\sqrt{2 \pi \, \sigma_Y^2}}e^{-\frac{Y^2}{2 \sigma_Y^2}},
\end{eqnarray}
with $\sigma_X^2, \sigma_Y^2$ the corresponding variances, which are defined as:
\begin{eqnarray}
&& \sigma_X^2 = \sigma_Y^2 \nonumber \\
&& = \frac{1}{2 \pi} \int_{-\infty}^{+\infty} |\chi(\omega)|^2 S_{F_X}(\omega)\, d \omega \nonumber \\
&& = \frac{k_B T}{k_0}, 
\end{eqnarray}
in which we introduced:
\begin{eqnarray}
\chi(\omega) & = & \frac{\Omega_m}{2k_0} \, \frac{1}{\Gamma_m/2 + \mathrm{i} \, \omega} ,\\
S_{F_X}(\omega) & = & 4 m_0 \Gamma_m \, k_B T ,
\end{eqnarray}
the mechanical susceptibility $\chi(\omega)$ in frequency-space and the force noise spectrum $S_{F_X}(\omega)$.
The variances are finite, as they should.

Eq. (\ref{energyEm}) tells us that energy is the sum of two uncorrelated squared Gaussian variables: this is known as a $(\chi_2)^2$ law. It results in an {\it exponential} distribution function:
\begin{equation}
    p(E_m)=\frac{1}{\sigma_E} e^{-\frac{E_m}{\sigma_E}} \, \Theta(E_m) , \label{boltz}
\end{equation}
which verifies $\langle E_m \rangle = \sigma_E$ and $\langle E_m^2 \rangle = 2 \sigma_E^2$, leading to an energy variance of $\sigma_E^2$. Since $\langle E_m \rangle= k_0 (\sigma_X^2 + \sigma_Y^2)/2$, one infers immediately that $\sigma_E=k_B T$ which matches the well-known equipartition result.\\

Eq. (\ref{boltz}) is nothing but the classical version of the Boltzmann  distribution. The final message is then that an exponential energy distribution {\it is equivalent} to a Gaussian motion distribution; with the magnitudes of the associated white spectra  related to $T$, the bath characteristic temperature. 

\section{Effective damping $\Gamma_{ef\!f} $}
\label{gammaoptic}

The microwave setup calibration is discussed in Appendix \ref{twpagain} below; on the other hand, the optomechanics coupling $G$ requires a specific measurement that we present here.
It is based on the exploitation of the mean sideband peak characteristics (averaging together all the data measured during the period $\Delta T$, using the 'blue detuned' pumping).
This is done for all acquisition speeds $\delta t$, except the fastest one (red curves in Fig. \ref{figbig}) which distorts the measured line; an effect commented in Appendix \ref{fastrescale}.
The Lorentzian fit enables to extract area (leading to mean energy, Fig. \ref{population}), peak position and half-height-width $\Gamma_{ef\!f}$. The latter can be fit to 
Eqs. (\ref{gammaeff},\ref{gammaopt}) as a function of injected power $P_{in}$, for each temperature $T$. This is shown in Fig. \ref{fig_damping}.

The slopes of these lines define the opto-mechanical coupling $G/(2 \pi) \approx 1.8 \cdot 10^{13} ~$Hz/m, and the $P_{in} \rightarrow 0$ extrapolation gives us the mechanical damping rate $\Gamma_m$. It is found to be linear in temperature in this range (see inset), in accordance with ref. 
\cite{XinPRAppl}.

 The same calibration can be done in 'red detuned' pumping, with a change of sign in the slope. 
Note that these slopes are independent of $T$, as they should be since the coupling is a pure geometrical effect.
The scatter in Fig. \ref{fig_damping} is genuine, and comes from the fluctuations of mechanical parameters (see Section \ref{oneovf}). 
Finally, the measurement of $\Gamma_{ef\!f}$
enables to compute the opto-mechanical gain %optomechanical gain?
${\cal G}$ for any $T$, $P_{in}$ (and any of the two 'red' or 'blue detuned' schemes).

\begin{figure}
        \includegraphics[width=\linewidth]{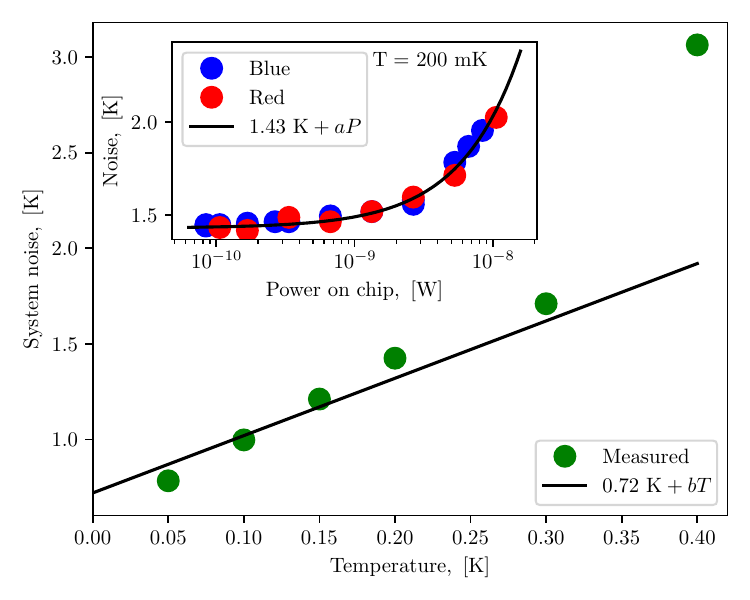}
        \caption{{\bf Main:} Microwave noise amplitude around $6~$GHz (referenced to the input of the TWPA amplifier), for low injection powers $P_{in}$.
        %%% Corr Luca new graph
        The line is a linear guide showing a temperature-dependent contribution when the mixing chamber plate temperature is regulated, and the $T \rightarrow 0$ limit of $0.8\pm0.1~$K (3 photons).
        Note the fast increase in noise above 0.35$~$K (see text).
        \newline
        {\bf Inset:} 
        Noise dependence to power $P_{in}$ measured at 200$~$mK, for the two different pumping schemes. The line is a linear guide to the eye, demonstrating the impact of input noise (see text).}
        \label{fig_TWPAnoise}
\end{figure}

\section{TWPA characterisation}
\label{twpagain}

Prior to the run, the microwave setup is characterized. 
%%%% corr add Nico
The gains and losses are measured carefully using a Vector Network Analyser (VNA).
For the injection lines, room temperature noise is suppressed by 50$~$dB attenuation affixed at different stages of the cryostat (zigzag blocks in Fig. \ref{scheme}). % I think its just one block there corresponding to all of them

On the detection side, the cryogenic HEMT provides a gain of 40$~$dB while the room temperature one has 30$~$dB. The microwave noise background around 6$~$GHz, referenced to the input of the cryo-HEMT, is then about 2.5$~$K. This is what is obtained from measurements performed before the mixer, with a spectrum analyser; when using the mixing technique, we obtain twice this background, as explained in the core of the manuscript. % I guess here we can add a bit about double side mixing
%%%% corr add Nico
Besides the calibration of our (passive) microwave elements, the HEMT noise figure has been verified, and validated, in a run using the {\it cold/hot load technique}: comparing the measured noise generated by a 50$~\Omega$ termination located on the 3$~$K plate to the one of a similar load bolted onto the mixing chamber stage (at 10$~$mK). A microwave switch mounted on the same mixing chamber plate enabled to connect one or the other loads while keeping the cryostat cold. 

At the lowest temperatures, the TWPA on/off %%%% add Luca
gain around 6$~$GHz is 10$~$dB, and the insertion loss of 'TWPA+directional coupler+circulator' is about 4$~$dB. 
This is not optimal since we work on the higher side of the bandpass of the amplifier, which was designed for a 15$~$dB gain at slightly lower frequencies.
%%%% corr Luca
Optimising the center frequency and the insertion losses, we infer that we could potentially win about $5 - 10 ~$dB  \cite{luca}. % I think it's more like 10dB
The gain degrades quickly as we increase the temperature of the parametric amplifier above 300$~$mK; besides, the intrinsic noise of the TWPA itself increases dramatically above this limit (because of free quasi-particles thermally excited in the aluminum layer, see Fig. \ref{fig_TWPAnoise}). Typically, it can still be used at 400$~$mK with a marginal gain, and essentially ruins the measurement chain above. % should we mention T_c/3 of Al as a cross-point

The measured noise of the full setup with the TWPA amplifier 'on' is shown in Fig. \ref{fig_TWPAnoise} (the three amplifier  %%%% corr add Nico
gains are taken into account, the noise being referenced to the input of the TWPA).
%%%% corr Luca
As we increase the mixing chamber plate temperature $T$, we see a thermal increase %of about $b T$
%%% corr Nico
in this noise 
%%% corr Luca , {\it without} correcting for the TWPA insertion loss
which is due to all the components present on the mixing chamber plate: our 50$~\Omega$ load, the microwave components (circulators, couplers) {\it and} 
the TWPA itself, see Fig. \ref{scheme}.
As such, modeling properly the $T$-dependent losses becomes quite involved, and is far outside of the scope of our manuscript. We therefore simply linearly fit the data in Fig. \ref{fig_TWPAnoise} in order to extract a reasonable estimate of our $T \rightarrow 0$ 
background noise.
The extrapolation slightly under-estimates the real limiting noise figure, since for $k_B T/(\hbar \omega_c) \ll 1$
it should flatten-out below typically 100$~$mK.
We therefore retain  a conservative value of about 0.8$~$K ($\pm 0.1~$K) at $T=0~$K, corresponding thus to about 3$~$photons, which matches expectations \cite{luca}.
%%% corr Luca
%Technically, the data presented in Fig. \ref{fig_TWPAnoise} has been shifted vertically to preserve the $T$-dependence of the TWPA, but still present the proper zero temperature limit; it thus corresponds to what would have been measured with an {\it ideal amplifier} presenting no insertion loss.
%%% I'm really not sure Nico will like it... We should be prepared to change it.
%%% Also, please have a careful look at the mixing business, maybe we can be more exact without over-complexifying the business?

\begin{figure}
        \includegraphics[width=\linewidth]{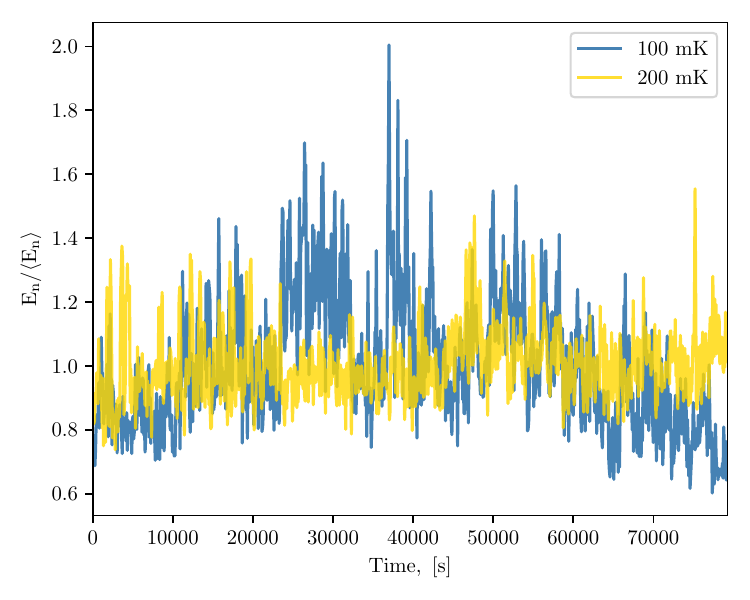}
        \caption{ %{\bf Main:} 
        Slow time-tracks normalised to mean for two temperatures, at microwave power around $7~$nW (see legend). 
        %%%XXXX Correct power??
        The colder set exhibits much {\it larger} excursions than the hotter one (see discussion in text).
        %\newline
        %{\bf Insert:} 
        }
        \label{fig_spikes}
\end{figure}

In the inset of Fig. \ref{fig_TWPAnoise}, 
we show the input power dependence of the measured noise: the background level increases with $P_{in}$, similarly for the two pumping schemes ('red' and 'blue detuned'). 
By measuring the cavity spectrum, we observe that this is actually due to an out-of-equilibrium photon population that increases with the pump power
\cite{XinPRAppl}. 
Classically, we model it with the voltage noise amplitudes $V_{P,\pm 1}$
appearing in Eq. (\ref{measure}), the sign $\pm$ referring to 'red' or 'blue detuned'. By using different microwave sources, or adding a notch filter, we can demonstrate that this noise is directly related to the quality of the pump signal. % (and {\it not} to a physical heating of an element of the circuitry). 
The actual power dependence seems to depend on the setup; it looks reasonably linear in Fig. \ref{fig_TWPAnoise} which is different from the findings of Ref. \cite{XinPRAppl}.
Besides, within our resolution it seems that the so-called technical heating of Fig. \ref{population}, and the $1/f$ contribution to the spectrum of Fig. \ref{figbig}, {\it do not} depend on this pump noise.
%%% This is true, right? It is right within our precision, which comes from our Anapico runs, but with anapico data was just bad.

\section{'Spikes' instability}
\label{spikes}

In Ref. \cite{XinPRAppl}, an instability in the dynamics of the beam was reported for temperatures lower than about 150$~$mK. 
It is visible as large amplitude peaks appearing in the sideband spectrum, which were nicknamed 'spikes'. 
The origin of this phenomenon is still unknown, and we presume that it should also impact our measurements. To which extent is the point of the present Appendix.

In Fig. \ref{fig_spikes} we plot two slow time-tracks taken at slightly different temperatures: 100$~$mK and 200$~$mK. 
Strikingly, we see that the amplitude of fluctuations is {\it much larger} on the colder data set, which is a clear signature of 'spikes'.

Let us comment in more details the data. When computing the mean energy $\langle E_n \rangle$, we see that the value extrapolated at zero injected power is actually {\it smaller} than expected, see Inset of Fig. \ref{population}.
On the other hand, the relative importance of technical heating grows as we cool down; if this is not taken into account properly (which is particularly difficult at low temperatures without TWPA), the inferred mode energy at a finite power $P_{in}$ would then be much {\it larger} than the thermodynamic value.

Interestingly, while the $1/f$ contribution to energy fluctuations increases as we cool down (Fig. \ref{1ovf}), the flat part still seems to reproduce very well the thermodynamic value, even at 100$~$mK (see Fig. \ref{spectlevel}).
It is thus very tempting to suggest that 'spikes' are inherently linked to the $1/f$ fluctuations, whatever might be the microscopic mechanism behind this.
How to understand a smaller mean energy compared to the thermodynamic temperature $T$ remains also an open question.

\begin{figure}
        \includegraphics[width=\linewidth]{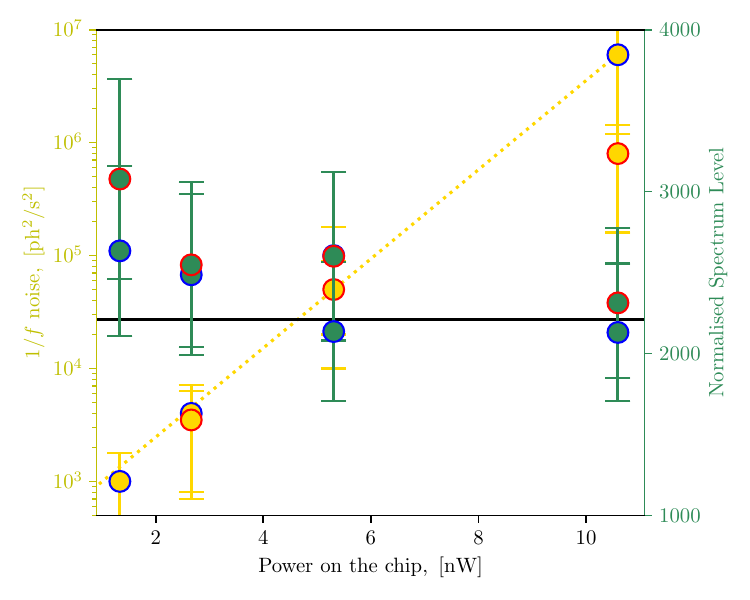}
        \caption{
        %{\bf Main:} 
        %\newline
        %{\bf Insert:} 
        %%% Maybe we should change the filling of the 1/f dots to keep only the outside blue. It could be yellowish? to match the other graph?
        Spectral contributions as a function of pump power $P_{in}$ measured at 200$~$mK with the two pumping schemes: 'red detuned' (red circled dots) or 'blue detuned' (blue circled).
        Left axis (log. scale): $1/f$ contribution $A_f$, in photons$^2$/s$^2$ (the dashed line is the Exponential fit of Fig. \ref{1ovf}). Right axis (lin. scale): Recalculated normalised phonon spectrum level (same parameter as in Fig. \ref{spectlevel}, no units). The horizontal line is obtained from the fit value of Fig. \ref{figbig}. 
        }
        \label{fig_redbluecomp}
\end{figure}%% Yes, defintiely. Give me 15 mins... Yes...
% I checked everything, maybe its better to discuss it on the phone? When? Great, Ill grab a coffee and wait for your call

\section{'Red' and 'blue' pumping schemes comparison}
\label{compareRB}

In order to validate our analysis, we compare 'red' and 'blue detuned' schemes. For the mean mechanical energy, this is done in Fig. \ref{population}. 
We see that indeed, correcting for the opto-mechanical gain ${\cal G}$ produces the same result.
In this Appendix, we shall concentrate on fluctuations.

In Fig. \ref{fig_redbluecomp} we plot the normalised phonon spectrum level 
$ \sqrt{\frac{\Gamma_m \,S_0 }{ {\cal G}^2}} \big / \left(\frac{\kappa_{ext}/2}{\kappa_{tot}} |\Gamma_{opt}|\right)$, recalculated from the photon flux spectrum fit (right axis, 200$~$mK data). 
We present both 'blue detuned' data (which can be found also in Fig. \ref{spectlevel}), and 'red detuned'. As for the mean mechanical energy, the agreement between the two pumping methods is very good; the difference being obviously that it is not possible to follow the fastest tracks with a 'red detuned' sheme, since we need the opto-mechanical gain ${\cal G}>1$ to do so.

In the same Fig. \ref{fig_redbluecomp}, we also show the corresponding 
$1/f$ components (left axis). However, we kept them in units of photons because transforming the graph into phonons following the same procedure as for the flat spectrum $S_0$ does not produce a much better plot [the fit of $A_f$ in Eq. (\ref{fitfunc}) is not that good].
Determining whether the origin of this effect is in the mechanics or the optics remains thus an open question.
However, we clearly see that 'red' and 'blue detuned' pumping data follow the same trend as a function of $P_{in}$: the $1/f$ term grows very quickly with increasing power (see Exponential fit). % We cannot distinguish them within our resolution
%%%% Yep, OK.

\section{Fastest tracks fit}
\label{fastfit}

The most difficult measurements are obviously the ones realised at the fastest acquisition speed.
%%%% add
For the slower tracks, the peaks are sufficiently well defined that the Lorentz fit error is small compared to the reproducibility; this is not true anymore for the fastest tracks, where both an increased error (discussed in this Appendix) and a bias (presented in the following Appendix G) exist.
%%%%
When opening the bandwidth, the background noise increases as well,  
and we can resolve the sideband peak only at the highest powers $P_{in}$, with the largest gains ${\cal G}$. Besides, 
with an acquisition bandwidth $1/\delta t$ larger than the peak width $\Gamma_{ef\!f}$, we lose information on {\it the shape} of the sideband: the imprint of the motion is visible as only 1-3 points higher than the background, see right top Inset in Fig. \ref{figbig}. 

The fitting procedure is then as follows: we first average together all the data sets taken over the period $\Delta T=1000 \, \delta t$. This produces a sideband peak which looks reasonably Lorentzian, with a width essentially given by the acquisition bandwidth.
Then, in the fitting routine that infers the sideband parameters of each datafile, we {\it fix} the Lorentz peak width to its mean value, and {\it constrain} the peak position to be around 0$~$Hz within only a few frequency-steps $1/\delta t$. % From measurements at slower acquisition rates we know that Gamma eff fluctuates as 1/f, so it is far to assume that it wont fluctuate within our acquisition rate and bandwidth
As such, the fit peak frequency position distribution looks like a (centered) truncated Gaussian, and our main fitting parameter is the height of the peak, or equivalently its area. 
Obviously, acquiring data faster than $\Gamma_{ef\!f}$
{\it should impact quantitatively} the extracted area value: this point is explicitly discussed in the following 
Appendix \ref{fastrescale}.
The opto-mechanics gain ${\cal G}$ is then computed from the known power (and temperature) dependence of $\Gamma_{ef\!f}$; note that at these acquisition speeds, $1/f$ noise in damping $\Gamma_m$ is irrelevant.

\begin{figure}
        \includegraphics[width=\linewidth]{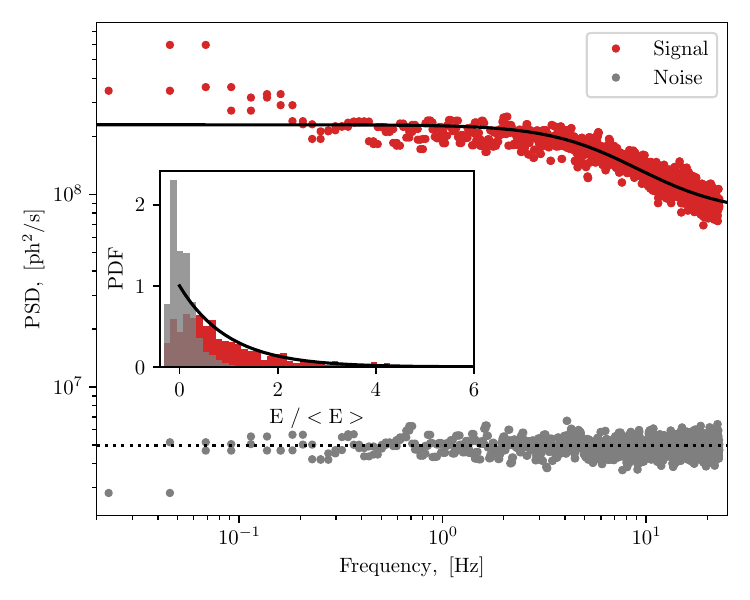}
        \caption{{\bf Main:} 200$~$mK fastest track Power Spectral Densities for peak (red data), and background (grey data). The full line corresponds to Eq. (\ref{fitfunc}) without the $1/f$ term, but with a constant noise contribution attributed to fitting noise (which has been subtracted in Fig. \ref{figbig}). The dashed line marks the level of background noise (see text for details).
        \newline
        {\bf Inset:} 
        Corresponding Probability Distribution Functions. The line is the exponential curve (see text).}
        \label{fig_spectralnoise}
\end{figure}

The great capability of this technique is that we can easily separate what is genuinely characteristic of the sideband, from what is  simply due to the noise background.
This is illustrated in Fig. \ref{fig_spectralnoise}, where we compare the Power Spectral Density calculated from the previous fitting, and the one obtained when constraining the fit position of the Lorentz peak far from the central value of $0~$Hz.
In the latter case, the obtained spectrum is white, and more than one order of magnitude smaller than what is obtained when fitting on the sideband (main graph). The Probability Distribution Function is centered on $0$, and clearly distinct from the exponential tail obtained with the sideband peak data (inset).
%%%% add
Note the slight negativity which comes from a non-constrained fit that also captures the background noise when no  signal is to be seen; this has been truncated in Fig. \ref{figbig} for clarity.

Finally, only the sideband processed data show the cutoff at $\Gamma_{ef\!f}$ in the spectrum; but the computed Power Spectral Density does not fall to zero above this value (see fit in Fig. \ref{fig_spectralnoise}).
This is presumably due to the fit error, which is distinct from the background noise (which is subtracted in Fig. \ref{figbig} for clarity); see following Appendix \ref{fastrescale} for details of fast-tracks fit corrections. 
%%%% add
As a matter of fact, the final scatter in Fig. \ref{spectlevel} corresponds to our ability of fitting the flat region of the FFTs of the type of Fig. \ref{figbig} (main graph), obtained for different temperatures. This finite error bar (which can be understood as our capability of defining $k_B$) corresponds here to 100 spectra averaged together, producing the typical scatter seen in Figs. \ref{figbig} and \ref{fig_spectralnoise}.

 % maybe we should mention that we subtract this constant from final plot in the main paper? As in caption we say to check text for details, and there is none)
% It's done

%%%% Heu... It's really N^2 in NumPy for the FFT??? It`s the same in the origin, it just does one but default
%%% Heu, one but default??? What do you mean?
%maybe we should merge two appendixes? F and G? they are very similar
%%% Good point, I had the same feeling. But in fact they do not exactly talk about the same aspect of the fast fitting... So, I dunno. I propose to check if it works like that. If not, then yes we could merge them.

\begin{figure}
        \includegraphics[width=\linewidth]{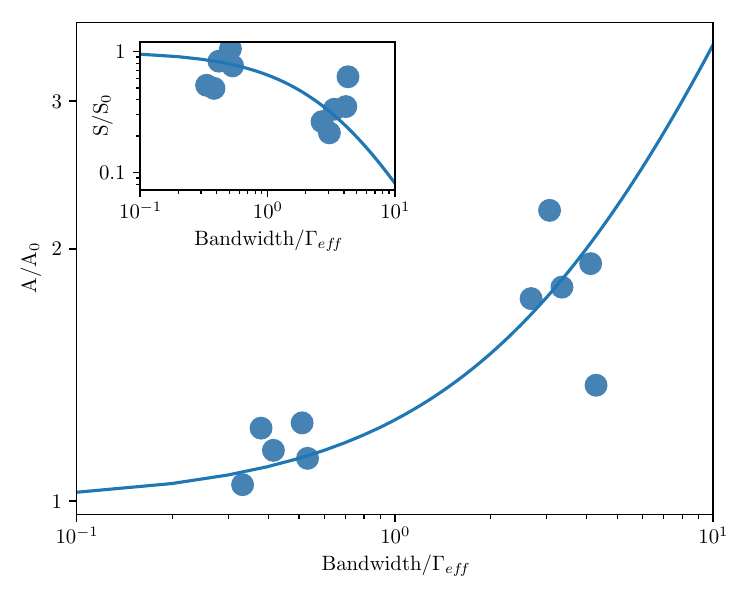}
        \caption{{\bf Main:} Mean area of sideband peak as a function of acquisition bandwidth $1/\delta t$ (only the two fastest settings), in normalized units ($A_0$ being the value corresponding to the slowest track; data taken at different powers and different temperatures). The line is a simple fit function: $f(x)=1+x/4$.
        \newline
        {\bf Inset:} 
        Corresponding normalised flat spectrum level as a function of bandwidth ($S_0$ being the reference value for slower acquisition speeds).
        %Have you seen my comment about the red/blue compare plot? What do you think?
        %%% We miss also one power in the time-tracks of spikes...
        %%% Do you confirm that the data for the effect of speed are measured at max power for different temperatures?? Or I got it wrong???
        % I didnt understand the last question
        The fit is: $1/f(x)^2$.}
        \label{fig_areacor}
\end{figure}

\section{Fast acquisition rate corrections}
\label{fastrescale}

At the fastest acquisition rate, we saw in the preceding Appendix that the shape of the sideband peak is altered: it becomes {\it broader}, its width being defined by the sampling rate. This is not the only impact of the fast-tracking. Comparing the mean area  obtained at different speeds $\delta t$, we also find out that it is {\it over-estimated}. On the contrary, when stitching the fastest spectrum to the others, we realise that we 
{\it under-estimate} fluctuations. This is summarized in Fig. \ref{fig_areacor}, in a universal plot with normalised axes.

The $x-$axis corresponds to the sampling bandwidth normalised to $\Gamma_{ef\!f}$. The $y-$axis is the mean area normalised to its value obtained at slow acquisitions (main), or the fit plateau in the fluctuation spectrum normalised to the value extracted with slow tracking (inset). These can be fit by very simple empirical laws, see Caption of Fig. \ref{fig_areacor}.

In practice, with our settings only the fastest tracks ($\delta t = 22~$ms, Fig. \ref{figbig}) need a rescaling. Note that it does not impact the fit of the plateau $S_0$ in Power Spectral Density, which is very clearly defined by slower acquisition rates. It is only needed for display purposes, when plotting the full-range data from $1/f$ to cutoff $\Gamma_{ef\!f}$. 
%%%% Added!!
The fitting routine itself also impacts  Fig. \ref{fig_areacor}, and its inherent bias is contained therein within our empirical dependencies. For a more profound analysis of fit biases, please see Ref. \cite{Extra}.

\section{Sliding averaging}
\label{sliding}

A first attempt to produce energy fluctuations spectra
had been made in Ref. \cite{Dylan}, with measurements performed down to the quantum regime. % should we clarify what is sliding average? Or just say, that it is explained in ref?
While the idea was clearly defined, the resolution of the experiment was very far from the requirements needed to produce the results we describe here. 
To obtain fittable data, the Authors had to process a 'sliding averaging' on the acquired measurements (averaging together $\# n$ neighboring files, while shifting this window through the whole set of data); and the extracted spectrum characteristics did {\it not} present the expected thermodynamic behavior.
The nature of these slow fluctuations remained thus an open question in this publication.

\begin{figure}[t]
        \includegraphics[width=\linewidth]{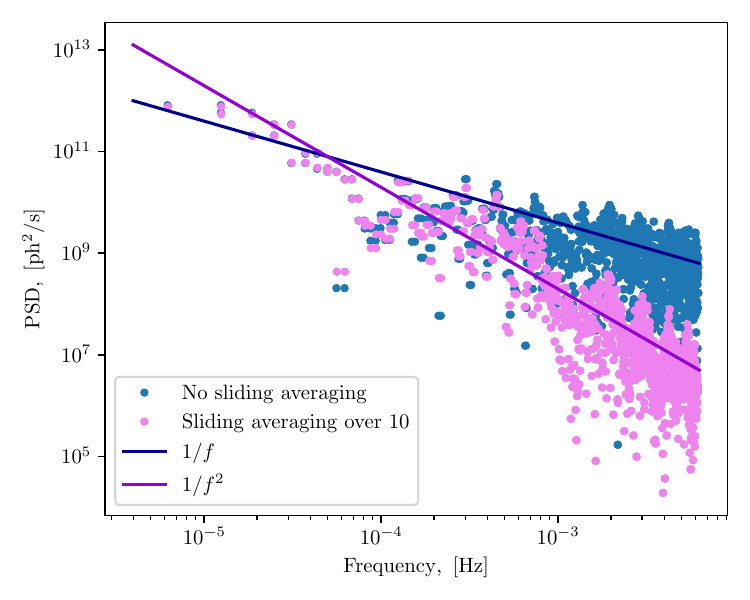}
        \caption{%{\bf Main:} 
        %\newline
        %{\bf Insert:} 
        Power Spectral Density obtained from data acquired at 200$~$mK (slowest rate, same set as Fig. \ref{figbig}).
        We present the original spectrum, and the one obtained when applying a 'sliding average' on the measurement. Lines are fits (see text).
        }
        \label{fig_Slidingaver}
\end{figure}

In the present Appendix we  study the effect of 'sliding averaging' on our own data. In 
Fig. \ref{fig_Slidingaver}, we plot the Power Spectral Density obtained with our raw slow data measured at 200$~$mK, together with the one obtained when processing a 'sliding averaging' (with a $\# n=10$ file averaging window). 
We clearly see that the averaging acts as a {\it low pass filter}, which transforms the initial $1/f$ component of our data into a $1/f^2$ (see fit lines); it  completely suppresses the thermodynamic plateau $S_0$. 

We therefore conclude that 'sliding averaging' essentially preserves {\it only} the $1/f$ component of energy fluctuations. There is thus no particular information in the shape of the spectra obtained this way, since they are characteristic only of the filtering method. 
Especially, what looked like a very low frequency cutoff with a plateau is nothing but an artifact of filtering + ${\cal FFT}$ method. 
However, the $\sigma_E \propto \sqrt{T}$
law observed in Ref. \cite{Dylan} contains genuine information, which is characteristic of the (unknown) mechanism causing these slow fluctuations.

\newpage
\bibliography{apssamp}% Produces the bibliography via BibTeX.

\end{document}